\documentclass[aps,showpacs,superscriptaddress]{revtex4}
\usepackage[dvips]{graphicx}
\usepackage{epsfig,rotating}
\usepackage{amsmath,amssymb}
\numberwithin{equation}{section}

\newcommand{\be}{\begin{equation}}
\newcommand{\ee}{\end{equation}}
\newcommand{\bea}{\begin{eqnarray}}
\newcommand{\eea}{\end{eqnarray}}
\begin{document}
\title{Are direct photons a clean signal of a thermalized quark gluon plasma?}
\author{D. Boyanovsky}
\email{boyan@pitt.edu} \affiliation{Department of Physics and
Astronomy, University of Pittsburgh, Pittsburgh, Pennsylvania
15260, USA}\affiliation{LPTHE, Universit\'e Pierre et Marie Curie
(Paris VI) et Denis Diderot (Paris VII), Tour 16, 1er. \'etage, 4,
Place Jussieu, 75252 Paris, Cedex 05, France}
\author{H. J. de Vega}
\email{devega@lpthe.jussieu.fr} \affiliation{LPTHE, Universit\'e
Pierre et Marie Curie (Paris VI) et Denis Diderot (Paris VII),
Tour 16, 1er. \'etage, 4, Place Jussieu, 75252 Paris, Cedex 05,
France}\affiliation{Department of Physics and Astronomy,
University of Pittsburgh, Pittsburgh, Pennsylvania 15260, USA}

\date{\today}

\begin{abstract}
Direct photon production from a quark gluon plasma (QGP) in
thermal equilibrium is studied directly in real time. In contrast
to the usual S-matrix calculations, the real time approach is
valid for a QGP that formed and reached LTE a short time after a
collision and of  finite lifetime ($\sim 10-20~\mathrm{fm}/c$ as
expected at RHIC or LHC). We point out that during such finite QGP
lifetime the spectrum of emitted photons carries information on
the initial state. There is an inherent ambiguity in separating
the virtual from the observable photons during the transient
evolution of the QGP. We propose a real time formulation to
extract the photon yield which includes the initial stage of
formation of the QGP parametrized by an effective time scale of
formation $\Gamma^{-1}$. This formulation coincides with the
S-matrix approach in the infinite lifetime limit. It allows to
separate the virtual cloud as well as the observable photons
emitted during the pre-equilibrium stage from the yield during the
QGP lifetime. We find that the lowest order contribution
$\mathcal{O}(\alpha_{em})$ which does \emph{not} contribute to the
S-matrix approach, is of the same order of or larger than  the
S-matrix contribution during the lifetime of the QGP for a typical
formation time $\sim 1~\mathrm{fm}/c$. The yield for momenta
$\gtrsim 3 ~\mathrm{Gev}/c$ features a power law fall-off $\sim
T^3 \Gamma^2/k^{5}$ and is larger than that obtained with the
S-matrix for momenta $\geq 4~\mathrm{Gev}/c$. We provide a
comprehensive numerical comparison between the real time and
S-matrix yields and study the dynamics of the build-up of the
photon cloud and the different contributions to the radiative
energy loss. The reliability of the current estimates on photon
emission as well as theoretical uncertainties on the details of
the initial state are discussed.

\end{abstract}

\pacs{11.10.Wx,12.38.Bx,12.38.Mh,13.85.Qk}
 \maketitle

%\tableofcontents

\section{ Introduction}\label{sec:intro}
Amongst the different potential experimental signatures of the
formation and evolution of a quark gluon plasma (QGP) that is
conjectured to be formed in ultrarelativistic heavy ion
collisions, hard electromagnetic probes, namely  direct photons
and dileptons are considered to be very
promising\cite{feinberg,mclerran}. Photons and dilepton pairs only
interact electromagnetically and their mean free paths are much
larger than the size of the QGP, thus these electromagnetic probes
leave the hot and dense region  after formation without further
scattering, carrying with them clean information of the early
stages of the collision.  Therefore a substantial effort has been
devoted to obtaining a theoretical assessment of the spectra of
direct photons and dileptons emitted from a thermalized
QGP\cite{feinberg}-\cite{renk}. Preliminary assessments concluded
that direct photon emission from a thermalized QGP can  be larger
than that from the hadronized phase\cite{kapusta,gale}, sparking
an intense effort to obtain reliable estimates of the direct
photon spectrum\cite{baier,aurenche,AMY}. For recent reviews on
theoretical and phenomenological aspects of electromagnetic
probes, see\cite{alam,revphoton,revphoton2}.

The first observation of direct photon production in
ultrarelativistic heavy ion collisions has been reported by the
 WA98 collaboration in
${}^{208}\mathrm{Pb}+{}^{208}\mathrm{Pb}$ collisions at
$\sqrt{s}=158~\mathrm{Gev}$ at the Super Proton Synchrotron (SPS)
at CERN\cite{WA98}. The results display a clear {\bf excess} of
direct photons above the expected background from hadronic decays
in the range of transverse momentum $p_T > 1.5 ~\mathrm{Gev}/c$ in
the most central collisions. These results provide an experimental
confirmation of the feasibility of  direct photons as reliable
probes in ultrarelativistic heavy ion collisions and may pave the
way for understanding the formation and evolution of a QGP.

A variety of fits of theoretical results to the experimental data
had been reported\cite{revphoton}, however,  the results seem
inconclusive, models with or without QGP emission seem to fit the
data in a manner compatible with models based solely on hadronic
`cocktails' (for a detailed review see\cite{revphoton}).

The current ultrarelativistic heavy ion program at the
Relativistic Heavy Ion Collider (RHIC-BNL) and the proposed heavy
ion program Alice at the forthcoming Large Hadron Collider
(LHC-CERN) have as a main goal to continue the experimental
pursuit of the long-sought QGP with beam energies of $\sqrt{s}
\sim 200 ~\mathrm{A Gev}$ for $\mathrm{Au}+\mathrm{Au}$  at RHIC
and up to $\sqrt{s}\sim 5500 \mathrm{A Gev}$ for
$\mathrm{Pb}+\mathrm{Pb}$ at CERN. This active experimental
program with the possibility of statistical analysis on
event-by-event basis justifies the theoretical assessment of
experimental probes at a deeper level. For a recent summary of
measurements at RHIC see\cite{jacobs}.

The S-matrix approach to calculating the photon yield from a QGP
in local thermal equilibrium treats the plasma as stationary and
with an infinite lifetime, while it is clear that QGP is a
transient, non-equilibrium state\cite{geiger,FOPI}. Current
theoretical understanding suggests that a QGP may be formed  $\sim
1~\mathrm{fm}/c$ after a nucleus-nucleus collision and thermalizes
via parton-parton scattering. The subsequent evolution is assumed
to be described by hydrodynamic expansion until the temperature
cools down to the hadronization scale $\sim 160~\mathrm{Mev}$. At
RHIC the initial temperature of the plasma is expected to be of
the order of  $300~\mathrm{Mev}$, and assuming Bjorken's
longitudinal expansion with a cooling law $T(t) = T_i
(t_i/t)^{\frac{1}{3}}$ it is expected that the lifetime of the QGP
is of order $\lesssim 10~\mathrm{fm}/c$ for a hadronization
temperature of about $160~\mathrm{Mev}$. At the LHC the initial
temperature is expected to reach $\sim 450~\mathrm{Mev}$ and the
lifetime of the QGP would be expected to be of order $\sim
20-30~\mathrm{fm}/c$. The transverse size of the QGP formed in the
most central collisions is of the order of the radius of the
nucleus which for $\mathrm{Pb}+\mathrm{Pb}$ is about
$7~\mathrm{fm}$ thus the typical space-time dimension of QGP in
local thermal equilibrium is about 10 fm.

Despite the fact that the quark gluon plasma, if formed, will
occupy a finite and rather small volume in space time, the
S-matrix approach to obtain the photon and dilepton yields treats
the plasma as  a medium in thermal equilibrium and  of infinite
extent in space-time\cite{geiger,FOPI}. The production rate
obtained from this approach is then input into a spacetime
evolution combined with a hydrodynamic expansion of the
plasma\cite{alam,revphoton,revphoton2}. A recent analysis of
photon production along these lines\cite{renk} to fit the data
from WA98\cite{WA98} suggests that the large $p_T$ region is
dominated by the first few fm/c of (hydrodynamic) evolution and is
very sensitive to the early stages of the evolution.

The issue of a finite space-time extension of the QGP and the
hadronic phase has received attention with respect to the emission
of photons and dileptons. The influence of a finite \emph{spatial}
size of the plasma has been addressed for the emission of thermal
photons\cite{wong,sarkar} and more recently for  thermal
dileptons\cite{ruuskanen2} from a hadronic gas, where the breaking
of detailed energy-momentum conservation by finite size effects
was studied.

Preliminary studies of the finite \emph{lifetime} effects on the
photoproduction yield were  reported in ref.\cite{boyanphoton}.
The results of those studies pointed out the importance of
non-equilibrium real time processes whose contribution is
subleading in the infinite lifetime limit, but that are of the
same order or larger than the S-matrix contribution during the
lifetime of a QGP expected at RHIC and LHC. Two main consequences
of the study in refs.\cite{boyanphoton} are:

i) during a finite lifetime the spectrum of direct photons is
sensitive to the initial conditions that lead to a thermalized QGP
with the large $p_T$ region of the spectrum more sensitive to the
initial stages, and

ii) to lowest order $\alpha_{em}$ the spectrum resulting from the
non-equilibrium processes flattens for momenta $p_T >
2~\mathrm{Gev}/c$. The sensitivity of the large $p_T$ part of the
spectrum to initial conditions has also been pointed out in
ref.\cite{renk}, and perhaps coincidentally, the WA98
data\cite{WA98} displays a flattening of the spectrum for $p_T
\geq 1.5 ~\mathrm{Gev}/c$.

\bigskip

{\bf Goals of this article:} The goals of this article are to
continue the study of direct photon production from a QGP in local
thermodynamic equilibrium with a finite lifetime,  directly in
real time. We focus on the following aspects:
\begin{itemize}
\item{Assessing the contribution to the direct photon spectrum
from the \emph{lowest order} processes that are subleading in the
infinite time limit. These processes are $q\bar{q} \rightarrow
\gamma$ and $q \rightarrow q\gamma$ and correspond to the one loop
contribution to the photon polarization, namely of order
$\alpha_{em}$. The contributions of these processes vanish in the
infinite time limit and do not contribute to the rate obtained
from the S-matrix approach, but do contribute to the yield during
a finite lifetime. Focusing on the lowest order contributions we
identify the dynamical aspects of photon production in real time
in the simplest possible case. This study  highlights that there
are contributions to \emph{all orders} in $\alpha_s$ that are
being missed by the S-matrix calculation.}

\item{A detailed analysis of the dynamics of the build-up of the
virtual photon cloud and to provide a systematic effective
description of the initial stage between the collision and
thermalization that allows a clear separation of the virtual
photons. We discuss the inherent difficulties associated with an
unambiguous identification of the virtual photon cloud during a
finite time interval.}

\item{A systematic description of direct photon production during
a finite time interval including the initial preparation of the
state. }

\item{An analytic and numerical comparison of the real time yield
obtained in lowest order, namely of $\mathcal{O}(\alpha_{em})$ and
the S-matrix yield, of order $ \alpha_s \; \alpha_{em}\;
\ln\frac{1}{\alpha_s} $\cite{aurenche,AMY} to assess the potential
experimental significance of the processes that are missed by the
S-matrix calculation but that contribute to the direct photon
yield from a QGP with a finite lifetime. We provide a
comprehensive numerical study of the direct photon yield to lowest
order $\mathcal{O}(\alpha_{em})$ with an analysis of the
spectrum.}

\item{ A study of the dependence of the  spectrum on the initial
conditions prior to the onset of local thermal equilibrium. This
study  reveals important aspects of the initial conditions prior
to thermalization that influence the spectrum.}

 \item{ A study of the radiative energy loss, in particular the
 contributions associated with the interaction energy as well as
 the cooling of the plasma by photon emission.}

%new addition #1
 \item{ A simple energy-time uncertainty argument would suggest that
 for momenta
 larger than the inverse lifetime of the QGP, the effects of a finite lifetime
 should be subleading. Our  study clearly shows this expectation
 \emph{not} to bear out. In fact we show that contributions from the
 region $\omega\neq k$
 in the imaginary part of the photon polarization are very important
 during the finite
 lifetime and of the same order (or larger ) than the usual result
 valid solely for $\omega=k$ even for photons with large transverse momentum.}
%end of new addition #1.

\end{itemize}

This article is organized as follows: in section \ref{sec:Smatrix}
we revisit the S-matrix approach to highlight its caveats. In
section \ref{sec:realtime} we present the real time formulation to
photon production beginning with a full gauge invariant treatment
of the electromagnetic interaction of quarks. In section
\ref{sec:photonyield} we  provide a simple and transparent
derivation of the expression for the photon production yield in
real time to lowest order in $\alpha_{em}$ and finite QGP
lifetime. This formulation reproduces the results obtained in
ref.\cite{boyanphoton} by a more general kinetic description and
is explicitly shown to coincide with the S-matrix formulation in
the infinite QGP lifetime limit. In this section we address the
issue of initial conditions and in particular the subtle but
important aspects associated with the formation of the photon
cloud. In this section we present a detailed analysis of the
radiative energy loss and the different contributions, providing
an analytic and numerical study of the total energy radiated
during the lifetime of the QGP. In section \ref{sec:adiabatic} we
address the issue of the electromagnetic dressing of the initial
state (density matrix) by providing an initial density matrix that
includes the photon cloud parametrized by a formation time scale
of the QGP after the parton stage following an ultrarelativistic
heavy ion collision, conjectured to be $\sim 1~\mathrm{fm}/c$.
This parametrization interpolates smoothly between the adiabatic
preparation of asymptotic states and the uncorrelated initial
state assumed in the S-matrix calculation. The consideration of
such initial states (density matrix) allows us to address the
issue of the formation time and includes in a phenomenological
manner the photon cloud of the pre-equilibrium stage.

Our conclusions are presented in section \ref{sec:conclusions}.

\section{S-matrix approach and its caveats}\label{sec:Smatrix}

In order to highlight the shortcomings of the S-matrix  approach
to calculate photon emission,  and to establish contact with the
real-time approach to photon production introduced in Sec.~IV, we
now summarize some important aspects of the scenario of QGP
formation and evolution and   the S-matrix approach to the
calculation of photon emission.

As mentioned in the introduction QGP is conjectured to be formed
in ultrarelativistic heavy ion collisions from the deconfinement
of strongly interacting quarks and gluons in the incoming nuclei.
The details of the dynamics of the collision are not completely
understood, nor, in particular, the electromagnetic aspects of the
parton distribution functions. It is conjectured that immediately
after the collision the partons are almost free and parton-parton
scattering leads to a state of (local) thermal equilibrium on a
time scale $\sim 1~\mathrm{fm}/c$ \emph{after} the collision. The
photons emitted during the pre-equilibrium stage are assumed to
leave the medium\cite{srivageiger}.  The thermalized QGP undergoes
adiabatic hydrodynamic expansion during a lifetime of $\sim 10-20~
\mathrm{fm}/c$ after which the plasma hadronizes. The QGP in local
thermal equilibrium under the strong interactions is not in
equilibrium under the electromagnetic interactions resulting in
photons emitted directly from the thermalized plasma.

The S-matrix approach to the calculation of photon emission begins
by writing the Hamiltonian  in the form
\begin{equation}
H = H_0+H_{\rm int},\quad H_0 = H_{\rm QCD}+H_{\gamma} \; , \quad
H_{\rm int} = e\int d^3x\,J^{\mu} A_{\mu} \; ,\label{totalH}
\end{equation}
where $H_{\rm QCD}$ is the full QCD Hamiltonian, $H_{\gamma}$ is
the free photon Hamiltonian, and $H_{\rm int}$ is the interaction
Hamiltonian between quarks and photons with $J^{\mu}$ the quark
electromagnetic current, $A^{\mu}$ the photon field, and $e$ the
electromagnetic coupling constant.

Consider that at some initial time $t_i$ the state $|i\rangle$ is
an eigenstate of $H_0$ with no photons. The transition amplitude
at time $t_f$ to a final state $|f,\gamma_{\lambda}({\vec
p})\rangle\equiv |f\rangle\otimes|\gamma_{\lambda}({\vec
p})\rangle$, again an eigenstate of $H_0$ but with one photon of
momentum ${\vec p}$ and polarization $\lambda$, is up to an
overall phase given by
\begin{equation}
 S(t_f,t_i) = \langle f,\gamma_{\lambda}({\vec
 p})|U(t_f,t_i)|i\rangle \; ,\label{transamp}
\end{equation}
where $U(t_f,t_i)$ is the time evolution operator in the
interaction representation
\begin{eqnarray}
U(t_f,t_i) &=& {\mathrm T}\,\exp\left[-i\int_{t_i}^{t_f}H_{{\rm
int},I}(t) \; ,dt\right]\nonumber\\
&=& 1-ie\int_{t_i}^{t_f}dt\int d^3x ~J^{\mu}_{I}({\vec x},t)
~A_{\mu,I}({\vec x},t)+\mathcal{O}(e^2) \; ,\label{umatx}
\end{eqnarray}
where the subscript $I$ stands for the interaction representation
in terms of $H_0$. In the above expression we have approximated
$U(t_f,t_i)$ to first order in $e$, since we are interested in
obtaining the probability of photon production to lowest order in
the electromagnetic interaction. The usual $S$-matrix element for
the transition is obtained from the transition amplitude
$S(t_f,t_i)$ above in the limits $t_i \rightarrow -\infty$ and
$t_f \rightarrow +\infty$
\begin{equation}
S_{fi}= S(+\infty,-\infty) =-\frac{ie}{\sqrt{2E}}\int d^3x
\int^{+\infty}_{-\infty}dt ~ e^{iP^\mu x_\mu}~
\varepsilon^{\lambda}_\mu ~\langle
f|J^\mu(x)|i\rangle+\mathcal{O}(e^2) \; ,\label{S}
\end{equation}
where $E=|{\vec p}|$ and $P^\mu=(E,{\vec p})$ are the energy and
four-momentum of the photon, respectively, and
$\varepsilon^{\lambda}_\mu$ is its polarization four-vector. Since
the states $|i\rangle$ and $ |f\rangle$ are eigenstates of the
{\em full} QCD Hamiltonian $H_{\rm QCD}$, the above $S$-matrix
element is obtained {\em to lowest order} in the electromagnetic
interaction, but {\em to all orders} in the strong interaction. We
note that the $S$-matrix element in effect is the amplitude for
the transition between asymptotic states $|i;{\rm
in}\rangle\rightarrow|f,\gamma_{\lambda}({\vec p});{\rm
out}\rangle$, i.e., $S_{fi}=\langle f,\gamma_{\lambda}({\vec
p});{\rm out}|i;{\rm in}\rangle$, where $|f,\gamma_{\lambda}({\vec
p});{\rm out}\rangle\equiv|f;{\rm
out}\rangle\otimes|\gamma_{\lambda}(\vec p);{\rm out}\rangle$.
Here, $|\gamma_{\lambda}({\vec p});{\rm out}\rangle$ is the
asymptotic {\sl out} state with one photon of polarization
$\lambda$ and momentum ${\vec p}$, and $|i;{\rm in}\rangle$
($|f;{\rm out}\rangle$) is the asymptotic {\sl in} ({\sl out})
state of the quarks and gluons.

The rate of photon production per unit volume from a QGP in
thermal equilibrium at temperature $T$ is obtained by squaring the
$S$-matrix element, summing over the final states, and averaging
over the initial states with the thermal weight $e^{-\beta
E_i}/Z(\beta)$, where $\beta=1/T$, $E_i$ is the eigenvalue of
$H_0$ corresponding to the eigenstate $|i\rangle$, and
$Z(\beta)=\sum_i e^{-\beta E_i}$ is the partition function. Using
the resolution of identity $1=\sum_f |f\rangle\langle f|$, the sum
of final states leads to the electromagnetic current correlation
function. Upon using the translational invariance of this
correlation function, the two space-time integrals lead to
energy-momentum conservation multiplied by the space-time volume
$\Omega=V(t_f-t_i)$ from the product of Dirac delta functions. The
term $t_f-t_i\rightarrow +\infty$ is the usual interpretation of
$2\pi\delta(0)$ in the square of the energy conserving delta
functions.

These steps lead to the following result for the photon production
rate in the S-matrix approach~\cite{mclerran,ruuskanen} \be
\frac{dN}{d^4x}=\frac{1}{\Omega}\frac{1}{Z(\beta)}\frac{d^3p}{(2\pi)^3}
\sum_{i,f,\lambda}e^{-\beta E_i}~|S_{fi}|^2  = -e^2\,
g^{\mu\nu}~W^<_{\mu\nu}(P)~ \frac{d^3p}{2E(2\pi)^3} \; ,\label{A}
\ee where $W^<_{\mu\nu}(K)$ is the Fourier transform of the
thermal expectation value of the current correlation function
defined by
\begin{equation}
W^<_{\mu\nu}(K) =\int d^4x\,e^{iK\cdot x}\,\langle
J_\mu(0)J_\nu(x)\rangle_\beta  \; .
\end{equation}
In the expression above $\langle\cdots\rangle_\beta$ denotes the
thermal expectation value. To lowest order in $e^2$ but to all
orders in the strong interactions, $W^<_{\mu\nu}(K)$ is related to
the retarded photon self-energy $\Pi^R_{\mu\nu}(K)$ by~\cite{gale}
\begin{equation}
e^2\,W^<_{\mu\nu}(K)=\frac{{\rm
Im}\Pi^R_{\mu\nu}(\omega=k,k)}{e^{k/T}-1} \; . \label{genfor}
\end{equation}
Thus, one obtains the (Lorentz boost) invariant photon production
rate
\begin{equation}
k\frac{dN}{d^3p\,d^4x}=-\frac{g^{\mu\nu}}{(2\pi)^3}\frac{{\rm
Im}\Pi^R_{\mu\nu}(\omega=k,k)}{e^{k/T}-1} \; . \label{invarate}
\end{equation}

All the calculations of the photon production yield from a
thermalized QGP in equilibrium begin by obtaining ${\rm
Im}\Pi^R_{\mu\nu}(\omega=k,k)$ to calculate the \emph{rate}. The
most recent result up to leading logarithmic order in the strong
coupling has been obtained in ref.\cite{AMY}.

We have reproduced the steps leading to Eq.~(\ref{invarate}),
which is the expression for the photon production rate used in all
S-matrix calculations in the literature, to highlight several
important steps in its derivation in order to compare and contrast
to the real-time analysis discussed below. The main features of
the above result that will be compared to the real time
computation are the following:

\begin{itemize}
\item{ The initial states $|i\rangle$ are averaged with the
thermal probability distribution at the initial time $t_i$ for
quarks and gluons. In the usual calculation this initial time
$t_i\rightarrow -\infty$, as emphasized above and the initial
state describes the photon vacuum and a thermal ensemble of quarks
and gluons. Thus the quarks and gluons are assumed to have
thermalized in the \emph{infinite past}. Furthermore, this
treatment also assumes that the quarks and gluons are
\emph{asymptotic} states in the infinite past, thus neglecting the
fact that these are confined in the colliding nuclei before the
collision.  }

\item { The transition amplitude is obtained via the time
evolution operator $U(t_f,t_i)$ evolved up to a time $t_f$ and the
transition amplitude is obtained by projecting onto a state
$|f\rangle$ at time $t_f$, which in the calculation is taken $t_f
\rightarrow +\infty$. The sum over the final states leads to the
electromagnetic current correlation function averaged over the
initial states with the Boltzmann probability distribution, i.e.,
the thermal expectation value of the current correlation
function.}

 \item { Taking $t_f\rightarrow +\infty$ and $t_i
\rightarrow -\infty$ and squaring the transition amplitude leads
to {\em energy conservation} and an overall factor $t_f-t_i$. The
rate (transition probability per unit time per unit volume $V$) is
finally obtained by dividing by $(t_f-t_i)V$. The important point
here is that taking the limit of $t_f-t_i \rightarrow +\infty$
results in {\em two} important aspects: energy conservation and an
overall factor of the time interval $t_f-t_i$. The resulting rate
is independent of the time interval and only depends on the photon
energy (and obviously the temperature).}

\end{itemize}

{\bf Main assumptions in the S-matrix approach:}. In order to
compare our methods and results with those obtained within the
usual S-matrix framework described above, it is important to
highlight the main assumptions that are implicit in {\em all}
previous calculations of photon production from a thermalized QGP
 and that are explicitly displayed by the derivation above.

\begin{itemize}

\item{ The  initial state at $t_i$ (which in the usual calculation
is taken to $-\infty$) is taken to be a thermal equilibrium
ensemble of quarks and gluons but the {\em vacuum state} for the
physical transverse photons. }

\item{ The usual calculation of the rate to lowest order in the
electromagnetic coupling, entails that there are no
electromagnetic corrections to the intermediate states, namely,
there is no photon \emph{dressing} of the states that enter in the
thermal density matrix.  }

\item{  Taking $t_i \rightarrow -\infty, t_f \rightarrow +\infty$
manifestly assumes that quarks and gluons are \emph{asymptotic
states} in the infinite past and in the infinite future. Obviously
this is inconsistent with the fact that before the collision
quarks and gluons should be described in terms of their parton
distribution functions in the nuclei. Furthermore assuming quarks
and gluons to be asymptotic states as $t_f \rightarrow +\infty$
manifestly ignores the hadronization phase transition to a
confined phase at a finite time of order $10-20 ~\mathrm{fm}/c$.
In ref.\cite{renk} confinement in the initial state had been
encoded in a `confinement factor', namely a phenomenological
parameter included to account for the effects of confinement. }

\item{ Assuming the QGP to have equilibrated at $t_i \rightarrow
-\infty$ and taking $t_f\rightarrow +\infty$ makes explicit that
the plasma is assumed to be described as a stationary state in
thermal equilibrium at all times. }

\item{ The buildup of population of photons is neglected along
with the electromagnetic dressing of quarks, these assumptions are
generally invoked to justify a calculation of the yield or rate to
lowest order in the electromagnetic coupling. }

\item{The rate obtained from a stationary state of thermal
equilibrium is then assumed to be valid in each fluid cell (of
spatial size larger than the mean free path) which is taken as the
local rest frame. The invariant rate (independent of time) is then
written in terms of the proper time and fluid rapidity by
performing a Lorentz boost and assuming that the temperature is a
function of the proper time. The resulting rate is then integrated
during the space-time history of the plasma in combination with a
hydrodynamic description of the expansion. During the
hadronization transition the yield is obtained from  a Maxwell
construction of the coexistence region (under the assumption of a
first order transition). The lever rule is invoked to obtain the
photon yield from the mixed phase. Thus despite the fact that the
rate has been obtained by taking the initial and final times to
$\mp \infty$ it is used to extract  the photon yield during a
\emph{finite} lifetime, and even during phase
coexistence\cite{alam,revphoton,revphoton2}. }

\end{itemize}

{\bf Caveats:}

\bigskip

The main reason that we delve on the specific steps of the usual
computation and on the detailed analysis of the main assumptions
is to emphasize the inconsistencies in applying this approach to
an expanding QGP of {\em finite lifetime}.

\begin{itemize}

\item{ Hydrodynamic evolution is an {\em initial value
problem}\cite{bjorken,blaizot}, namely,  the state of the system
is specified at an initial (proper) time surface to be of local
thermodynamic equilibrium at a given initial temperature,  and the
hydrodynamic equations are evolved in time to either the
hadronization or freeze-out surfaces if the equation of state is
available for the different stages. The calculation based on the
S-matrix approach takes the time interval to infinity, extracts a
{\em time-independent rate} treating the QGP as a stationary
state, and inputs this rate, assumed to be valid for every cell in
the comoving fluid, in the hydrodynamic evolution during a finite
lifetime.}

\item{ There is also a physical inconsistency in using the
S-matrix yield in a hydrodynamic evolution for very large photon
energy. A hydrodynamic description, which is based on local
thermodynamic equilibrium,  is valid on spatial scales  larger
than the mean free path for parton-parton collisions in the plasma
$\lambda \sim 0.5 ~\mathrm{fm}$. Thus photon momenta $k \geq 2-3$
 Gev probe distances shorter than the mean free
path, and most likely the contribution to the direct photon
spectrum for transverse momenta larger than about $2-3$ Gev cannot
be reliably extracted from a S-matrix calculation. Recent
measurements of elliptic flow at RHIC\cite{jacobs} suggest that
hydrodynamics is a reliable description up to $p_T \sim
2~\mathrm{Gev}/c$ but the data for $v_2(p_T)$ show large
departures from hydrodynamics (including pQCD)for $p_T >
2~\mathrm{Gev}/c$ for charged particles (minimum bias)\cite{v2pT}.

The low energy region of the photon spectrum is dominated by pion
decay and bremsstrahlung in the hadronic phase and after
freeze-out. Thus, the (transverse) momentum interval in which
direct photons could be reliable experimental probes of  a
thermalized QGP is $0.1~\mathrm{Gev} \lesssim k \lesssim
3~\mathrm{Gev}$. }

\item{ Yet another caveat is that despite the fact that the time
interval is taken to infinity, namely much larger than the photon
thermalization time scale, photons are assumed \emph{not to
thermalize and to leave the medium}. The buildup of the photon
population is {\em neglected} under the assumption that the mean
free path of the photons is larger than the size of the plasma and
the photons escape without rescattering. This assumption also
neglects the prompt photons produced during the pre-equilibrium
stage. Indeed, Srivastava and Geiger~\cite{srivageiger} have
studied direct photons from a pre-equilibrium stage via a parton
cascade model that includes pQCD parton cross sections and
electromagnetic branching processes. The usual computation of the
prompt photon yield during the stage of a {\em thermalized} QGP
assumes that these photons have left the system and the
computation is therefore carried out to lowest order in
$\alpha_{em}$ with an initial photon vacuum state, namely also
ignoring the virtual cloud of photons that dress the charged
particles in the plasma. As emphasized above, in taking the final
time $t_f$ to infinity in the $S$-matrix element the assumption is
that the thermalized state is {\em stationary}, while in
neglecting the buildup of the population the assumption is that
the photons leave the system without rescattering and the photon
population never builds up. These assumptions lead to considering
photon production only to the lowest order in $\alpha_{em}$, since
the buildup of the photon population will necessarily imply higher
order corrections. Although these main assumptions are seldom
spelled out in detail, they underlie all S-matrix calculations of
the photon production from a thermalized QGP.}

\item{All calculations of the rate based on the S-matrix approach,
obtain the imaginary part of the photon polarization to lowest
order in $\alpha_{em}$ and in a perturbative expansion in terms of
$\alpha_s$ (including leading logarithmic terms). This expansion
assumes that $\alpha_s$ is small but the coupling depends on the
temperature scale. While it could be argued that at the initial
temperatures expected to be achieved at RHIC and LHC $\alpha_s$
may be small, clearly the perturbative expansion breaks down in
the expanding scenario, when the temperature becomes near the
critical for hadronization $T_c \sim \Lambda_{QCD}\sim
160~\mathrm{Mev}$. Thus, the regime of validity of \emph{all}
S-matrix  calculations of the yield in a perturbative expansion in
$\alpha_s$ is actually limited by the \emph{lifetime} of the QGP.
Hence a perturbative evaluation of the rate must be understood to
be valid on a time scale of the order of or shorter than the
actual lifetime of the QGP, which then casts further doubts on the
infinite time limit.  }

\end{itemize}

As stated in the introduction, however, the QGP produced in
ultrarelativistic heavy ion collisions is intrinsically a
transient and nonequilibrium state. Since the spectrum of direct
photons is deemed to be a clean experimental  probe of the early
stages of evolution of a QGP, it is therefore of phenomenological
importance to study nonequilibrium effects on direct photon
production from an expanding QGP with a \emph{finite lifetime}
with the goal of establishing potential experimental signatures.

The current understanding of the QGP formation, equilibration, and
subsequent evolution through the quark-hadron (and chiral) phase
transitions is summarized as follows. A pre-equilibrium stage
dominated by parton-parton interactions and strong colored fields
which gives rise to quark and gluon production on time scales
$\lesssim 1$ fm/$c$~\cite{geiger}. The produced quarks and gluons
thermalize via elastic collisions on time scales $\sim 1$ fm/$c$.
Hydrodynamics is probably the most frequently used model to
describe the evolution of the next stage when quarks and gluons
are in local thermal equilibrium (although perhaps not in chemical
equilibrium)~\cite{bjorken,blaizot}. The hydrodynamical picture
assumes local thermal equilibrium (LTE), a fluid form of the
energy-momentum tensor and the existence of an equation of state
for the QGP. The subsequent evolution of the QGP is uniquely
determined by the hydrodynamical equations, which are formulated
as an {\em initial value problem} with the initial conditions
specified at the moment when the QGP reaches LTE, i.e., at an
initial time $t_i \sim 1$ fm/$c$. The (adiabatic) expansion and
cooling of the QGP is then followed to the transition temperature
at which the equation of state is matched to that describing the
mixed and hadronic
phases~\cite{ruuskanen,alam,sollfrank,srivastava}.

Our main observation is that the usual computations based on
S-matrix theory  extract a time independent rate after taking the
infinite time interval, which  is then used in a calculation of
the photon yield during a {\em finite time} hydrodynamic
evolution. There is a conceptual inconsistency in this approach,
which merits a detailed  study based on the real time evolution of
the photon distribution, which we undertake below.

\section{Real time approach}\label{sec:realtime}

\subsection{Gauge Invariance}\label{sec:GI}
Before we focus on the calculation of the photon yield in real
time, we address the  issue of abelian gauge invariance to
highlight that the results of the real time approach are fully
gauge invariant. Since the relevant interaction is
electromagnetic, we focus our discussion on the abelian gauge
coupling. Let us consider the following Lagrangian density for one
massless fermion species
\begin{equation}
\mathcal{L}= \bar{\psi}(i\!\not\!\partial -e \not\!\!A)\psi
-\frac{1}{4}F_{\mu \nu}F^{\mu \nu} \; ,\label{lagra}
\end{equation}
where the zero-temperature mass of the fermion $m$ has been
neglected since we consider the high temperature limit $T\gg m$.
We begin by casting our study directly in a manifestly gauge
invariant form.  In the abelian case it is straightforward to
reduce the Hilbert space to the gauge invariant states and to
define gauge invariant fields. This is best achieved within the
canonical Hamiltonian formulation in terms of primary and
secondary class constraints. In the abelian case there are two
first class constraints:
\begin{equation}
\pi_0=0 \; ,\quad {\bf
\nabla}\cdot\boldsymbol{\pi}=-e\,\psi^\dagger\psi \; ,
\label{firstclass}
\end{equation}
where $\pi_0$ and $\boldsymbol{\pi}=-\mathbf{E}$ are the canonical
momenta conjugate to $A^0$ and $\mathbf{A}$, respectively.
Physical states are those which are simultaneously annihilated by
the first class constraints and physical operators \emph{commute}
with the first class constraints. Writing the gauge field in terms
of transverse and longitudinal components as
$\mathbf{A}=\mathbf{A}_L+\mathbf{A}_T$ with ${\bf
\nabla}\times\mathbf{A}_L=0~;~ {\bf \nabla}\cdot\mathbf{A}_T=0$
and defining
\begin{equation}
\Psi(x)=\psi(x) \; e^{ie\int d^3y {\bf
\nabla}_\mathbf{x}G(\mathbf{x}-\mathbf{y})\cdot\mathbf{A}_L(y)} \;
,
\end{equation}
where $G(\mathbf{x}-\mathbf{y})$ the Coulomb Green's function
satisfying ${\bf \nabla}_{\mathbf{x}}^2
G(\mathbf{x}-\mathbf{y})=\delta^3(\mathbf{x}-\mathbf{y})$, after
some algebra using the canonical commutation relations one finds
that $\mathbf{A}_T(x)$ and $\Psi(x)$ are \emph{gauge invariant}
field operators.

The Hamiltonian can now be written solely in terms of these gauge
invariant operators and when acting on gauge invariant states the
resulting Hamiltonian is equivalent to that obtained in Coulomb
gauge. However we emphasize that we have \emph{not} fixed any
gauge, this treatment, originally introduced by Dirac is
manifestly gauge invariant. The instantaneous Coulomb interaction
can be traded for a gauge invariant Lagrange multiplier field
which we call $A^0$, leading to the following Lagrangian density
\be \mathcal{L} = \bar{\Psi}(i\!\not\!\partial -e\gamma^0 A^0 +
e\boldsymbol{\gamma}\cdot \mathbf{A}_T)\Psi+ \frac{1}{2}
\left[(\partial_{\mu} \mathbf{A}_T)^2+ ({\bf \nabla}
A^0)^2\right]\label{lagrainv} \; . \ee We emphasize that $A^0$
should \emph{not} be confused with the temporal gauge field
component.

In the gauge invariant sector of the Hilbert space, namely between
states annihilated by the first class constraints, and
generalizing to $N_f$ flavors of quarks,  the Hamiltonian is given
by,
\begin{equation}\label{GIH}
H=\int d^3 x\left[\frac{1}{2}\left(\vec{E}^2_T+\vec{B}^2\right)+
\Psi^{\dagger}(-i\vec{\alpha}\cdot\vec{\nabla})\Psi+e \;
\vec{J}\cdot \vec{A}_T\right]+H_{coul} \; ,
\end{equation}
\noindent where \be \vec{J} = \sum_{i=1}^{N_f}\frac{e_i}{e} \;
\bar{\Psi}_i \; \vec{\gamma}\; \Psi_i \label{currents} \; , \ee
\noindent is the \emph{gauge} invariant current and $H_{coul}$ is
the abelian Coulomb interaction which will be irrelevant for our
considerations. While this Hamiltonian is equivalent to that
obtained in Coulomb gauge we emphasize that we have not imposed
any gauge fixing, the Dirac procedure is manifestly gauge
invariant. In particular the interaction part of the Hamiltonian
that will be relevant for the discussion of photon production to
lowest order, namely
\begin{equation}
H_I= e\int d^3x ~\vec{J}\cdot{\vec A}_T
\end{equation}
is manifestly gauge invariant. Furthermore, states constructed out
of the non-interacting Fock vacuum by combinations of the
\emph{gauge invariant} operators
$\Psi^{\dagger},\Psi,\vec{A}_T,\vec{E}_T$ are obviously gauge
invariant.

This discussion makes explicit the gauge invariance of the real
time formulation.

Including now the non-abelian color interaction between quarks and
gluons the total Hamiltonian is given by
\begin{equation}\label{totalham}
H =  H_{QCD}[\Psi]+\int
d^3x~\frac{1}{2}~(\vec{E}^2_T+\vec{B}^2)+e\,\int d^3 x ~
 \mathbf{J}\cdot \mathbf{A}_T + H_{coul} \; ,
\end{equation}
\noindent where $H_{QCD}[\Psi]$ is the QCD Hamiltonian in absence
of electromagnetism but in terms of the gauge invariant (under
abelian gauge transformation) fermion field $\Psi$ and the
subscript $T$ refers to transverse components. We have extended
the fermion content to $N_f$ flavors and the charge of each flavor
species in units of the electron charge is included in the
corresponding current.

Expanding the gauge invariant field $\vec{A}_T$ in terms of
creation and annihilation operators in a volume $V$
\begin{equation}
\mathbf{A}_T(\vec{x})= \sum_{\mathbf{k},\lambda}
\frac{\mathbf{\epsilon}_{\mathbf{k},\lambda}}{\sqrt{2Vk}}
\left[a_{\mathbf{k},\lambda} \; e^{i\mathbf k\cdot
\mathbf{x}}+a^{\dagger}_{\mathbf{k},\lambda} \; e^{-i\mathbf
k\cdot \mathbf{x}} \right] \; ,
\end{equation}
\noindent with $\mathbf{\epsilon}_{\mathbf{k},\lambda}$ the usual
transverse polarization vectors, the \emph{gauge invariant} photon
number operator for polarization $\lambda$ is given by
\begin{equation}\label{numberop}
\hat{n}_{\mathbf{k},\lambda} =
a^{\dagger}_{\mathbf{k},\lambda}\,a_{\mathbf{k},\lambda} \; .
\end{equation}

\subsection{Time evolution}

Time evolution in quantum mechanics and quantum field theory is an
initial value problem. Given the total Hamiltonian $H$ the time
evolution of a density matrix is completely determined by
specifying the density matrix at an initial time $t_0$. Once the
initial density matrix is specified at the initial time, its time
evolution is completely determined by the unitary time evolution
operator $e^{-iH(t-t_0)}$. In the case under consideration the
Hamiltonian $H$ is \emph{time independent and gauge invariant} and
given by eq. (\ref{totalham}). The ensuing real time dynamics is
therefore completely specified by prescribing the initial density
matrix at the initial time $t_0$.

As discussed above the S-matrix calculation implicitly assumes
that the initial density matrix is that of thermal equilibrium for
quarks and gluons at an initial time $t_0 = -\infty$ and
explicitly assumes that initially there are no photons. With this
choice of initial conditions, since the initial state has been
prepared at $t_0 = -\infty$ there is no memory of the initial
state, in agreement with a stationary state in thermodynamic
equilibrium.

Our goal is to relax this assumption of a thermal stationary state
and study the consequences of a true real time evolution as befits
the physical problem of a thermalized QGP emerging about 1 $fm/c$
after a nucleus-nucleus collision and evolving during a finite
lifetime of about 10-20 $fm/c$  towards a hadronization (and
confinement) transition.

The real time evolution of the density matrix after the initial
time $t_0$ is given by
\begin{equation}
\hat{\rho}(t)= e^{-iH(t-t_0)} \; \hat{\rho}(t_0) \;
e^{iH(t-t_0)}\; .
\end{equation}
 The total number of photons of momentum $k$ per
unit volume at a given time $t$, namely the photon yield is given
by (assuming translational invariance)
\begin{equation}\label{photonnumber}
(2\pi)^3~\frac{d {N}(t)}{d^3x d^3k} = \sum_{\lambda}
\mathrm{Tr}\left[\hat{\rho}(t)\,\hat{n}_{\mathbf{k},\lambda}\right]\;,
\end{equation}
\noindent with $\hat{n}_{\mathbf{k},\lambda}=
\hat{a}^\dagger_{\mathbf{k},\lambda} \;
\hat{a}_{\mathbf{k},\lambda}$.  In order to compute the direct
photon yield to lowest order in the electromagnetic coupling, it
is convenient to write the total Hamiltonian given by
eq.(\ref{totalham}) as \bea
&&H =  H_0 + H_I \;,\label{totih}\\
&&H_0 = H_{QCD}[\Psi]+\int d^3x \; \frac{1}{2} \;
(\vec{E}^2_T+\vec{B}^2)\; ,  \; \label{H0} H_I  = \int d^3 x  \;
\mathbf{J}\cdot \mathbf{A}_T \;, \nonumber \eea \noindent where
the current $\mathbf{J}$ is given by eq. (\ref{currents}) and we
have neglected the Coulomb term since it will not contribute to
the direct photon yield to order $\alpha_{em}$.

The time evolution operator is given by, \be e^{-iH(t-t_0)} =
e^{-iH_0 t} \; U(t,t_0) \; e^{iH_0 t_0}\;, \label{timeevol} \ee
\noindent where $U(t,t_0)$ the unitary time evolution operator in
the interaction picture of $H_0$ given by \be U(t,t_0) =
1-i\int_{t_0}^{t} dt' \;  H_I(t')+\mathcal{O}(e^2) ~~;~~H_I(t)=
e^{iH_0t} \; H_I \; e^{-iH_0t} \label{timeevolop}\; , \ee
\noindent where we have only considered the lowest order in the
electromagnetic coupling.

It is then convenient to pass to the interaction picture of $H_0$
namely the full QCD Hamiltonian and \emph{free} electromagnetism
by defining the initial density matrix in the interaction picture
of $H_0$ as \be \hat{\rho}_{ip}(t_0) = e^{iH_0
t_0}\;\hat{\rho}(t_0)\;e^{-iH_0 t_0}\label{rhoiniip}\; . \ee For
the case of interest the initial density matrix describes a quark
gluon plasma in equilibrium under the strong interactions, it must
therefore commute with $H_0$ in which case
$\hat{\rho}_{ip}(t_0)=\hat{\rho}(t_0)$. At any given time $t$
\begin{equation}
\hat{\rho}(t) = e^{-iH_0t} \; U(t,t_0) \; \hat{\rho}_{ip}(t_0) \;
U^{-1}(t,t_0) \; e^{iH_0t}\; . \label{rhotip}
\end{equation}
Since the number operator $a^{\dagger}_{\mathbf{k},\lambda} \;
a_{\mathbf{k},\lambda}$ commutes with $H_0$ we find for the direct
photon yield at time $t$ the following exact expression
\begin{equation}\label{photonnumberoft}
(2\pi)^3~\frac{dN(t)}{d^3xd^3k} = \sum_{\lambda}
\mathrm{Tr}\left[U(t,t_0) \; \hat{\rho}_{ip}(t_0) \; U^{-1}(t,t_0)
\; \hat{n}_{\mathbf{k},\lambda}\right] \; .
\end{equation}
In the interaction picture of $H_0$ the time evolution of
$\mathbf{J}$ and $\mathbf{A}_T$ is given by
\begin{eqnarray}
\mathbf{J}(\vec{x},t') & = & e^{iH_{QCD}t'} \; \mathbf{J}(\vec{x})
\; e^{-iH_{QCD}t'}\; , \cr \cr \mathbf{A}_T(\vec{x},t') & = &
\sum_{\mathbf{k},\lambda}
\frac{\mathbf{\epsilon}_{\mathbf{k},\lambda}}{\sqrt{2Vk}}
\left[a_{\mathbf{k},\lambda} \; e^{i\mathbf k\cdot \mathbf{x}} \;
e^{-ikt'}+a^{\dagger}_{\mathbf{k},\lambda} \; e^{-i\mathbf k\cdot
\mathbf{x}} \;  e^{ikt'}\right]\;, \label{ATip}
\end{eqnarray}
\noindent where $\mathbf{\epsilon}_{\mathbf{k},\lambda}$ are the
transverse polarization vectors.

Given an initial density matrix, the photon yield can be
calculated by inserting a complete set of eigenstates of $H_0$ and
computing the required matrix elements.

We note that the real time expression for the photon yield
eq.(\ref{photonnumberoft}) is \emph{exact} in terms of the full
time independent Hamiltonian and is gauge invariant provided that
the density matrix is constructed with physical states. Evolution
of quantum states or density matrices is an initial value problem,
once the density matrix has been specified at an initial time, its
time evolution is completely determined by the Hamiltonian.

We carry out this program below to lowest order in the
electromagnetic coupling for cases that are relevant to the
description of direct photons from a QGP.

\section{Photon yield}\label{sec:photonyield}

We compute here the photon yield for a QGP with finite lifetime in
the real time approach.

\subsection{Thermalized QGP, no initial
photons:}\label{subsec:nophotons}

 We begin the study of the real
time dynamics of direct photon production by considering that at
an initial time $t_0 \sim 1~\mathrm{fm}/c$ the QGP is in thermal
equilibrium (under the strong interactions) and there are \emph{no
initial photons}. Such initial density matrix is compatible with
the usual assumption on the initial state invoked in the S-matrix
calculation described in section \ref{sec:Smatrix} but makes
explicit that such state describes an initial value problem from a
thermalization time $t_0$ taken to be of the order of
$1~\mathrm{fm}/c$.
\begin{equation}\label{rhoininophots}
\hat{\rho}(t_0)= \sum_{n_q}e^{-\beta E_{n_q}} \;
|n_q\rangle\langle n_q| \otimes |0_{\gamma}\rangle\langle
0_{\gamma}|~~;~~ H_{QCD}|n_q\rangle = E_{n_q}|n_q\rangle \; ,
\end{equation}
\noindent and $|0_{\gamma}\rangle$ is the photon vacuum,
annihilated by the \emph{gauge invariant operators}
$a_{\mathbf{k},\lambda}$. The incoming nuclei coasting along the
light cone are exact eigenstates of the full Hamiltonian H. Hence,
neglecting bremsstrahlung off the Coulomb field of the nuclei
there is no photon emission prior to the collision. Thus, the
state $|0_{\gamma}\rangle$ is the `in' state up to the time of the
collision, which is annihilated by the `in' operator
$a_{\mathbf{k},\lambda}$. Therefore the initial density matrix eq.
(\ref{rhoininophots}) is consistent with neglecting the photons
produced between the collision and the onset of thermalization,
namely during the pre-equilibrium stage.

We will study in detail alternative and more general initial
states that include photons and correlations in a later section
(see section \ref{sec:adiabatic} below).

Obviously the initial density matrix given by
eq.(\ref{rhoininophots}) is gauge invariant and commutes with
$H_0$ in eq.(\ref{H0}), it follows that
$\hat{\rho}(t_0)=\hat{\rho}_{ip}(t_0)$ and it describes a thermal
ensemble in equilibrium under the strong interactions with no
photons. This assumption  is compatible with \emph{all the
calculations} of direct photon production from an equilibrated QGP
available in the literature. Studying the time evolution of this
initial state allows us to address the dynamics of the formation
of the virtual photon cloud and to highlight the inherent
difficulty in separating the observable photons from those in the
virtual cloud in the plasma during a finite lifetime.

Defining as
$|n_q;m_{\gamma}\rangle=|n_q\rangle\otimes|m_{\gamma}\rangle$ the
eigenstates of $H_0$ with $m_{\gamma}$ photons (we do not specify
the wavevector and polarization to avoid cluttering of notation)
we can compute the matrix elements in eq.(\ref{photonnumberoft})
by inserting a complete set of these eigenstates. To lowest order
in the electromagnetic coupling the set of intermediate states
that contribute to the photon number $\langle
\hat{n}_{k,\lambda}\rangle$ contain only one photon of momentum
$k$ and polarization $\lambda$.

After a straightforward calculation we find, to lowest order in
the electromagnetic coupling
\begin{eqnarray}
(2\pi)^3\frac{d {N}(t)}{d^3x d^3k} & = &  \sum_{\lambda}
\frac{e^2}{2 k V} \int_{t_0}^t dt_1 \int_{t_0}^t dt_2 \int d^3x_1
\int d^3x_2 \; e^{-ik(t_2-t_1)} \;
e^{i\mathbf{k}\cdot(\mathbf{x}_2-\mathbf{x}_1)}
\times \nonumber \\
&  & \sum_{n_q,m_q}e^{-\beta E_{n_q}} \; \langle
n_q|\mathbf{\epsilon}_{\mathbf{k},\lambda}\cdot
\mathbf{J}(\vec{x_2},t_2)|m_q\rangle  \; \langle
m_q|\mathbf{\epsilon}_{\mathbf{k},\lambda}\cdot
\mathbf{J}(\vec{x_1},t_1)|n_q\rangle  \; . \label{currcurr}
\end{eqnarray}
Writing,
\begin{equation}
\mathbf{J}(\vec x,t)= e^{-i\vec{P}\cdot\vec{x}} \; e^{iH_{QCD}t}
\; \mathbf{J}(\vec{0},0) \; e^{-iH_{QCD}t} \;
e^{i\vec{P}\cdot\vec{x}} \; ,
\end{equation}
\noindent with $\vec{P}$ the momentum operator and choosing the
eigenstates $|n_q\rangle$ to be simultaneous eigenstates of
$\vec{P}$ with $\vec{P}|n_q\rangle = \vec{p}_{n_q}|n_q\rangle$, we
can write
\begin{equation}
\sum_{n_q,m_q}e^{-\beta E_{n_q}} \; \langle
n_q|\mathbf{\epsilon}_{\mathbf{k},\lambda}\cdot
\mathbf{J}(\vec{x_2},t_2)|m_q\rangle \; \langle
m_q|\mathbf{\epsilon}_{\mathbf{k},\lambda}\cdot
 \mathbf{J}(\vec{x_1},t_1)|n_q\rangle
= \int d^3p \; d\omega \; e^{-i\vec{p}\cdot(\vec{x}_2-\vec{x}_1)}
\; e^{i\omega(t_2-t_1)} \; \epsilon^i_{\mathbf{k},\lambda} \;
\epsilon^j_{\mathbf{k},\lambda} \;
 \sigma^>_{ij}(\vec{p},\omega) \; ,
\end{equation}
\noindent where
\begin{equation}
\sigma^>_{ij}(\vec{p},\omega)=\sum_{n_q,m_q} \; e^{-\beta E_{n_q}}
\; \langle n_q|{J_i}(\vec{0},0)|m_q\rangle  \; \langle
m_q|{J_j}(\vec{0},0)|n_q\rangle \;
\delta^3(\vec{p}-(\vec{p}_{n_q}-\vec{p}_{m_q})) \;
\delta(\omega-(E_{n_q}-E_{m_q})) \; . \label{specgreat}
\end{equation}
Carrying out the integrals in eq.(\ref{currcurr}) and summing over
the polarizations, we finally find,
\begin{equation}
 \frac{d {N}(t)}{d^3x d^3k} = \frac{e^2 }{k}  \int d\omega \,
\mathcal{P}^{ij}({\mathbf{k}}) \; \sigma^>_{ij}(\vec{k},\omega) \;
\frac{1-\cos[(\omega-k)(t-t_0)]}{(\omega-k)^2} \; , \label{yield1}
\end{equation}
\noindent where
\begin{equation}
\mathcal{P}^{ij}({\mathbf{k}})= \delta^{ij}-\frac{k^ik^j}{k^2}\; ,
\end{equation}
\noindent is the transverse projector.

In appendix \ref{appendix:piret} we show that
\begin{equation}
(2\pi)^3\, e^2\,\sigma^>_{ij}(\vec{k},\omega)=
\frac{n(\omega)}{\pi} \;
\mathrm{Im}\Pi_{ij}(\vec{k},\omega)~~;~~n(\omega)=\frac{1}{e^{\beta
\omega}-1} \; ,
\end{equation}
\noindent where $\mathrm{Im}\Pi_{ij}(\vec{k},\omega)$ is the
imaginary part of the retarded photon polarization tensor.
However,  the photon production yield is calculated to lowest
order in $\alpha_{em}$ and in principle to all orders in
$\alpha_s$.

Introducing the transverse photon polarization
\begin{equation}\label{transverse}
\Pi_T(k,\omega) \equiv \mathcal{P}^{ij}({\mathbf{k}})\,
\Pi_{ij}(\vec{k},\omega) \; ,
\end{equation}
\noindent we finally find the real time expression for the photon
yield to be given by
\begin{equation}
\frac{d N(t)}{d^3x d^3k} = \frac{1}{(2\pi)^3\,k}
\int_{-\infty}^{+\infty} \frac{d\omega}{\pi} \;
\frac{\mathrm{Im}\Pi_{T}(k,\omega)}{e^{\frac{\omega}{T}}-1} \;
\frac{1-\cos[(\omega-k)(t-t_0)]}{(\omega-k)^2}\label{yieldPi} \; ,
\end{equation}
\noindent where the photon polarization is obtained in lowest
order in $\alpha_{em}$ and in principle to all orders in
$\alpha_s$. This expression coincides with the lowest order result
obtained from a kinetic description in ref.\cite{boyanphoton}.

If the long time limit is taken and the following identity is used
 (see\cite{boyanphoton,boyankinetic})
\begin{equation}
\int_{-\infty}^{+\infty} d\omega \; F(\omega) \;
\frac{1-\cos[(\omega-k)(t-t_0)]}{(\omega-k)^2} \buildrel{t-t_0 \to
+\infty}\over=\pi (t-t_0) \; F(k) +\int_{-\infty}^{+\infty}d\omega
\; F(\omega) \; \mathcal{P} \frac{1}{(\omega-k)^2}
+\mathcal{O}\left(\frac{1}{t-t_0}\right) \label{tdelta} \; ,
\end{equation}
\noindent then \emph{if} $\mathrm{Im}\Pi_{T}(k,\omega=k)\neq 0$,
we obtain the long time limit of the total yield,
\begin{equation}
\frac{d N(t)}{d^3x d^3k} = \frac{1}{(2\pi)^3\,k} \left\{
\frac{\mathrm{Im}\Pi_{T}(k,\omega=k)}{e^{\frac{k}{T}}-1} \;
(t-t_0)+\int^{+\infty}_{-\infty}\frac{d\omega}{\pi}
\frac{\mathrm{Im}\Pi_{T}(k,\omega)}{e^{\frac{\omega}{T}}-1} \;
\mathcal{P}\frac{1}{(\omega-k)^2} \right\}\label{longtimeyield} \;
.
\end{equation}
>From the result (\ref{longtimeyield}) above, we obtain the long
time limit of the invariant rate,
\begin{equation}\label{invaratePiinfi}
k \frac{d N(t)}{d^4x d^3k}\buildrel{t-t_0 \to +\infty}\over=
\frac{1}{(2\pi)^3} \;
\frac{\mathrm{Im}\Pi_{T}(k,\omega=k)}{e^{\frac{k}{T}}-1}  \; .
\end{equation}
This is the same as the S-matrix result given by
eq.(\ref{invarate}), the expression (\ref{longtimeyield}) only
involves the \emph{transverse} part of the photon polarization, a
consequence of the manifestly gauge invariant treatment. Thus, we
highlight that in the \emph{infinite time limit} our result for
the invariant rate coincides with the usual one.

However,  taking the time to infinity introduces all the caveats
that were discussed in section \ref{sec:Smatrix} above. Thus, in
order to avoid these caveats, the initial time $t_0$ must be
interpreted as the time at which quarks and gluons thermalize. The
expression (\ref{yieldPi}) determines the total photon yield (per
unit volume) at a given time $t$, thus the total direct photon
yield at the hadronization time is obtained by setting
$t=t_{had}\sim 10-20~\mathrm{fm}/c$ for RHIC and LHC respectively.

The asymptotic long time expression (\ref{longtimeyield}),
features \emph{two} different terms. The first term, which grows
linearly in time leads to the \emph{rate} which is time
independent and is associated with the photons that are produced
per unit time per unit volume. The second, time independent term
can be interpreted as the total number of photons in the virtual
cloud dressing the quarks in the medium.

This  manifest separation between the time independent term that
describes the photon cloud associated with the charged particles,
and the photons produced at constant rate (namely the term in the
yield that grows linearly with $t-t_0$) \emph{only emerges} in the
long time limit.

For any \emph{finite} time interval there are contributions to the
photon yield from the whole range of $\omega \neq k$. The long
time asymptotic behavior  of the total photon yield
(\ref{yieldPi}) is determined by the behavior of
$\mathrm{Im}\Pi(\omega \sim k)$. In particular if
$\mathrm{Im}\Pi(k,\omega = k)=0$ there is no \emph{linear} time
dependence asymptotically, however if $\mathrm{Im}\Pi(\omega \sim
k) \propto (\omega-k)$ as is the case for the one loop
contribution to the polarization [of
$\mathcal{O}(\alpha_{em}\alpha_s)$, see below] then using the
formula\cite{boyanphoton,boyankinetic}
\begin{equation}\label{logi}
\int_0^{+\infty} dy\,\frac{p(y)}{y}\left[1-\cos(ykt)\right]
\buildrel{t \to +\infty}\over = p(0)\ln(kt) + \mathcal{O}(1) \; ,
\end{equation}
\noindent for $p(\infty) =0$. We thus conclude that the photon
yield will grow \emph{logarithmically in time} in this case.

The long time behavior of the yield is therefore determined by the
behavior of the $\mathrm{Im}\Pi(k,\omega)$ in the region $\omega
\sim k$. Generally,   the imaginary part of the photon
polarization for $\omega \sim k$ behaves as
\begin{equation}\label{threshold}
\mathrm{Im}\Pi_{T}(k,\omega)\buildrel{ \omega \to k}\over =
\mathrm{Im}\Pi_{T}(k,\omega=k)+ \mathrm{Im}\Pi'_T(k,\omega=k) \;
(\omega-k)+\mathcal{O}\left[(\omega-k)^2\right] \; ,
\end{equation}
\noindent then we find\cite{boyanphoton,boyankinetic} the long
time behavior of the yield to be generally given by
\begin{eqnarray}\label{genyield}
\frac{d N(t)}{d^3x d^3k} =  &&\frac{1}{(2\pi)^3\,k} \Bigg\{
\frac{\mathrm{Im}\Pi_{T}(k,\omega=k)}{e^{\frac{k}{T}}-1}(t-t_0)+
\frac{\mathrm{Im}\Pi'_T(k,\omega=k)}{\pi~(e^{\frac{k}{T}}-1)}~
\ln\left[k(t-t_0)\right]+ \nonumber \\ &&
 \int \frac{d\omega}{\pi} \;
\frac{\mathrm{Im}\Pi_{T}(k,\omega)}{e^{\frac{\omega}{T}}-1} \;
\mathcal{P}\frac{1}{(\omega-k)^2}+\mathcal{O}\left(\frac{1}{t-t_0}
\right) \Bigg\} \; ,
\end{eqnarray}
\noindent where the prime in the second term stands for a
derivative with respect to the frequency.  Obviously at very long
times the term with the linear time dependence will ultimately
dominate. However, for a finite time interval $(t-t_0)$, it is
possible that a lower order logarithmic term may give a comparable
contribution to or even be larger than a higher order  term linear
in time. Furthermore, it is clear from the above analysis that the
contributions to the imaginary part of the photon polarization for
$\omega \neq k$ will actually contribute to the yield during a
finite time interval. However, these contributions are absent in
the S-matrix  calculation which only extracts the linear time
dependence in the \emph{asymptotic long time limit} given by
eq.(\ref{invaratePiinfi}).

The main point of this discussion is to highlight the following
important issues:

\begin{itemize}
\item{The real time calculation reproduces the result of the
S-matrix approach in the asymptotic long time limit. Therefore the
usual S-matrix result is \emph{contained} in the real time
approach, which provides a detailed description of the process of
photon production during a finite lifetime. }

\item{During a finite time interval the region $\omega \neq k$ of
the imaginary part of the photon polarization contributes. The
contributions from different regions of the spectral density have
different time dependence. During a finite  time interval
contributions that are subleading in the asymptotic long time
limit can be of the same order as the term that becomes
asymptotically linear in time   and  which defines the rate. In a
finite time interval the contributions from the different regions
in the frequency integral cannot be separated. Thus, in order to
reliably understand the time dependence of the yield during a
finite time interval, one must find the imaginary part of the
photon polarization in the \emph{full range} of frequency, not
just at $\omega =k$ which only determines the asymptotic long time
behavior.  }

\item{ This discussion makes manifestly clear that the photon
\emph{yield} obtained from the S-matrix calculation of the rate,
namely extracting the linear time dependence in the asymptotic
long time limit, ignores  all other contributions which grow
slower in time but that do contribute to the yield for a finite
time interval. }
\end{itemize}

\subsection{Photon production in the hard thermal loop
approximation}

To make the above statement more quantitative and to begin our
study of the real time description of photon production within a
specific example  highlighting the conclusions above, we begin by
considering the imaginary part of the photon polarization tensor
in the hard thermal loop approximation\cite{brapis,lebellac}. This
approximation yields the leading result for the polarization
tensor for soft photons, namely $k \ll T$ and arises \emph{solely}
from in-medium processes. While we will obtain the full one-loop
contribution to the imaginary part of the photon self energy
below, the HTL limit \emph{only} features medium dependent
contributions. Thus the HTL limit allows us to address dynamical
issues solely associated with in-medium processes.

For two flavor of quarks (up and down, with three colors), the
imaginary  part of the transverse photon polarization in the hard
thermal loop approximation is given by\cite{boyanphoton}
\begin{equation}
\mathrm{Im}\Pi^{HTL}_{T}(k,\omega)= \frac{40 \pi^2 \alpha_{em}
T^2}{36} \frac{\omega}{k}\left(1-\frac{\omega^2}{k^2}
\right)\Theta(k^2-\omega^2) \label{HTL} \; .
\end{equation}
Obviously this contribution to the imaginary part vanishes
linearly as $\omega \rightarrow k$. However,  while
eq.(\ref{longtimeyield}) would lead to a constant, time
independent yield and therefore to a vanishing invariant rate
since $\mathrm{Im}\Pi_{ij}(\omega=k)=0$, the correct asymptotic
long time limit of the yield follows from eq.(\ref{genyield}) and
is given by\cite{boyanphoton}
\begin{equation}\label{HTLasy}
 \frac{d N^{HTL}(t)}{d^3x d^3k} \buildrel{t-t_0 \to
+\infty}\over = \frac{5\,\alpha_{em} T^2}{18 \pi^2\,k^2}\left[
n(k) \; \ln[k(t-t_0)] + \frac{1}{2k}\int^{k}_{-k} d\omega \;
\frac{\omega}{k} \;
\mathcal{P}\left[\frac{k+\omega}{k-\omega}\right] \;
\frac{1}{e^{\frac{\omega}{T}}-1}+
\mathcal{O}\left(\frac{1}{t-t_0}\right)\right] \;  .
\end{equation}
This expression would lead to an invariant rate that vanishes as
$\mathcal{O}(1/t) $ for $t\rightarrow +\infty$. However, for any
finite time there is a non-vanishing contribution to the yield.

The second, time independent  term in the eq.(\ref{HTLasy}) can be
identified with the virtual photon cloud \emph{in the thermal
bath}. This contribution is medium dependent and can only be
identified in the long time limit,  the separation between the
time independent and the time dependent contributions is
meaningful only in the long time limit. For any finite lifetime
there is no unambiguous separation between the different
contributions, and only the full photon number is meaningful.

A comparison with the photon equilibrium spectral density in the
HTL approximation\cite{lebellac} reveals that the integrand of the
time independent part in eq. (\ref{HTLasy}) is simply related to
the HTL wave function renormalization and the real part of the
photon polarization. Namely, the virtual cloud is actually
revealing the dynamics of formation of a plasmon quasiparticle in
the medium.

\begin{figure}[ht!]
\epsfig{file=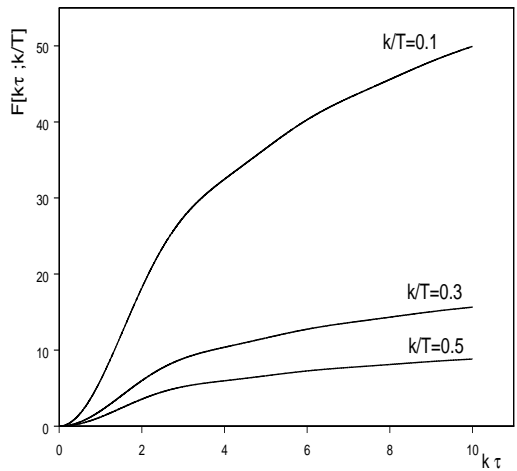,width=3in,height=3in}
 \caption{The
function $\mathcal{F}\left[k\tau;\frac{k}{T}\right]$ vs. $k\tau$
for $\frac{k}{T}=0.1;0.3;0.5$ respectively} \label{fig:htlyield}
\end{figure}

 The result eq.(\ref{HTLasy}) is valid in the asymptotic  long time
limit, for any arbitrary finite time we must use the full
expression given by eq.(\ref{yieldPi}). In particular, in the hard
thermal loop limit we find that the yield is given by
\begin{eqnarray}
\frac{d N^{HTL}(t)}{d^3x d^3k} & = & \frac{5}{36 \pi^2}
\frac{\alpha_{em} T^2}{k^2}
\mathcal{F}\left[k(t-t_0);\frac{k}{T}\right]  \; ,\cr \cr
\mathcal{F}\left[k\tau;\frac{k}{T}\right] & = & \int^1_{-1}dx\;
n\left[x\,\frac{k}{T}\right]x(1-x^2)
\frac{1-\cos\left[(x-1)k\tau\right]}{(x-1)^2}\label{yieldHTLoft}
\; .
\end{eqnarray}
The function $\mathcal{F}\left[k\tau;\frac{k}{T}\right]$ is
displayed in fig. \ref{fig:htlyield} for several values of the
ratio $k/T$. This figure  displays the logarithmic growth
determined by the asymptotic behavior (\ref{HTLasy}) at long
times.

\subsection{Full one loop polarization:}

As mentioned in the introduction, we will study the dynamics of
photon production during a finite lifetime by focusing on the
lowest order contribution to the photon polarization. This is a
quark one loop diagram   and is of order $\alpha_{em}$. The
imaginary part of this diagram vanishes at $\omega=k$, hence this
lowest order diagram does not contribute to the usual rate
obtained from the  long time limit as discussed in detail above.

The goal of this study is to understand the photon production
during the finite lifetime of the QGP from processes whose
contribution is subleading in the asymptotic long-time limit. Of
course diagrams which are higher order in $\alpha_s$  will also
give contributions to the photon production from the region
$\omega \neq k$, but focusing on the lowest order diagram we will
be able to extract important aspects of the dynamics that are
 missed by the S-matrix calculation and that could be experimentally relevant.

A comparison between real time yield obtained from the one loop
contribution to the photon polarization  and the S-matrix yield in
a wide  range of photon energy, in particular for large photon
momentum,  requires the full expression for the photon
polarization to one loop order.

A lengthy but straightforward calculation gives the following
result
\begin{equation}\label{totpi}
\mathrm{Im}\Pi^{T}(\omega,k)=\pi_0(\omega,k)+\pi_{LD}(\omega,k)+
\pi_{2P}(\omega,k)
\end{equation}
\noindent with $\pi_0;\pi_{2P}$ and $\pi_{LD}$ the zero
temperature, two fermion thermal cut and Landau damping
contributions, respectively, given by the following expressions
\begin{eqnarray}
\pi_0(\omega,k)  = && \frac{10}{9} \;
\alpha_{em}\,(\omega^2-k^2)\Theta(\omega^2-k^2)\,
\mathrm{sign}(\omega) \; , \label{pi0}\\
\pi_{2P}(\omega,k)  = && \frac{10}{3} \; \alpha_{em}\,T^2
\left(\frac{\omega^2}{k^2}-1\right)\Bigg[ \frac{k}{T} \ln
\left(\frac{1+e^{-W_+}}{1+e^{-W_-}} \right)-\frac{2T}{k}
\sum_{m=1}^{\infty}(-1)^{m+1}
\Big[\frac{2}{m^3}\left(e^{-m\,W_-}-e^{-m\,W_+} \right)\nonumber\\
&& -\frac{k}{T\,m^2}\left(e^{-m\,W_-}+e^{-m\,W_+} \right)
\Bigg]\Theta(\omega^2-k^2)\,\mathrm{sign}(\omega) \; ,  \label{pi2P} \\
 \pi_{LD}(\omega,k)  = && \frac{10}{3} \;
\alpha_{em}\,T^2 \left(1-\frac{\omega^2}{k^2}\right)\Bigg[
\frac{k}{T} \ln \left(\frac{1+e^{-W_-}}{1+e^{-W_+}}
\right)+\frac{2T}{k} \sum_{m=1}^{\infty}(-1)^{m+1}
\left(\frac{2}{m^3}+\frac{k}{T\,m^2} \right)\times\nonumber \\
&&  \left(e^{-m\,W_-}-e^{-m\,W_+} \right)
\Bigg]\Theta(k^2-\omega^2)\,\mathrm{sign}(\omega) \quad , \quad
W_{\pm} =  \Big | \frac{|\omega|\pm k}{2T}\Big | \label{piLD} \; .
\end{eqnarray}
The first two terms $\pi_0,\pi_{2P}$ arise from the process
$q\bar{q} \rightarrow \gamma$ and $\pi_{LD}$ from in-medium
bremsstrahlung $q\rightarrow \gamma q$. The long-wavelength limit
$k \ll T$ is dominated by $\pi_{LD}$ and simplifies to the HTL
expression eq. (\ref{HTL}).

\subsection{Dynamics of the virtual photon
cloud}\label{subsec:cloud}

As discussed above, the asymptotic long time limit is determined
by the behavior of $\mathrm{Im}\Pi_T(k, \omega)$ for $\omega \sim
k$, therefore it is convenient to separate the contribution from
the positive and negative frequency regions in the integral in eq.
(\ref{yieldPi}). Using the properties
$\mathrm{Im}\Pi_T(k,-\omega)=-\mathrm{Im}\Pi_T(k,\omega)$ and
$n(-\omega)=-[1+n(\omega)]$ we write
\begin{equation}
\frac{d N(t)}{d^3x d^3k}  =  \frac{d N^{(+)}(t)}{d^3x
d^3k}+\frac{d N^{(-)}(t)}{d^3x d^3k}\label{yieldtot}
\end{equation}
\noindent in terms of the  positive $(+)$ and negative $(-)$
frequency contributions respectively, given by
\begin{eqnarray}
\frac{d N^{(+)}(t)}{d^3x d^3k}  & =  & \frac{1}{(2\pi)^3\,k}
\int_{0}^{\infty} \frac{d\omega}{\pi}\; n(\omega)\;
\mathrm{Im}\Pi_{T}(k,\omega) \;
\frac{1-\cos[(\omega-k)(t-t_0)]}{(\omega-k)^2}\label{yieldPiposi}\\
\frac{d N^{(-)}(t)}{d^3x d^3k}  & = & \frac{1}{(2\pi)^3\,k}
\int_{0}^{\infty} \frac{d\omega}{\pi}\;[1+n(\omega)]\;
\mathrm{Im}\Pi_{T}(k,\omega) \;
\frac{1-\cos[(\omega+k)(t-t_0)]}{(\omega+k)^2}\label{yieldPinega}
\end{eqnarray}
For further analysis, it is convenient  to add together the terms
that feature the Bose-Einstein distribution function, thus we
write instead,
\begin{equation}
\frac{d N}{d^3x d^3k}   =  \frac{d N^{(T)}}{d^3x d^3k}+ \frac{d
N^{(V)}}{d^3x d^3k}\label{Nsplit} \; ,
\end{equation}
\noindent with
\begin{eqnarray}
\frac{d N^{(T)}(t)}{d^3x d^3k}  & =  & \frac{1}{(2\pi)^3\,k}
\int_{0}^{\infty} \frac{d\omega}{\pi}\; n(\omega)\;
\mathrm{Im}\Pi_{T}(k,\omega)\left\{
\frac{1-\cos[(\omega-k)(t-t_0)]}{(\omega-k)^2}+
\frac{1-\cos[(\omega+k)(t-t_0)]}{(\omega+k)^2}\right\} \; ,
\label{yielda}\\
\frac{d N^{(V)}(t)}{d^3x d^3k}  & = & \frac{1}{(2\pi)^3\,k}
\int_{0}^{\infty} \frac{d\omega}{\pi}\;
\mathrm{Im}\Pi_{T}(k,\omega)\;
\frac{1-\cos[(\omega+k)(t-t_0)]}{(\omega+k)^2}\label{yieldb}\; .
\end{eqnarray}
Both terms are positive, however,  while the frequency integral in
$\frac{d N^{(T)}(t)}{d^3x d^3k}$ is finite because of the
Bose-Einstein distribution function, the frequency integral in
$\frac{d N^{(V)}(t)}{d^3x d^3k}$  features divergences associated
with the virtual photon cloud, which can be seen as follows.

\begin{figure}[htbp]
\begin{center}
\epsfig{file=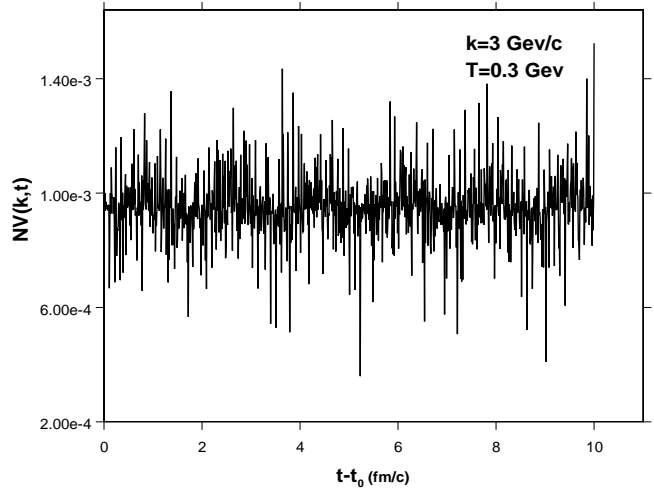,width=4in,height=3in}
 \caption{The  contribution $\frac{d N^{(V)}(t)}{d^3x d^3k}$  given by
 eq. (\ref{yieldb})  integrated up to
 $\omega_c=100~\mathrm{Gev}$ for $k=3~\mathrm{Gev/c};
 T=0.3~\mathrm{Gev}$ .} \label{fig:vircloud}
 \end{center}
\end{figure}

The  contribution $\frac{d N^{(V)}(t)}{d^3x d^3k}$  does not
feature resonant denominators, therefore it  remains positive and
does not grow in time at long time. The oscillatory terms  average
out on a short time interval, as shown explicitly in fig.
\ref{fig:vircloud} which displays the negative frequency
contribution integrated up to a frequency cutoff $\omega_c = 100
~\mathrm{Gev}$.

 Thus, after a very short transient time the average of $\frac{d
  N^{(V)}(t)}{d^3x d^3k}$ is obtained by neglecting the oscillatory cosine term
in eq. (\ref{yieldb}).

The remaining time independent term in eq. (\ref{yieldb}) features
two distinct contributions:

i) the zero temperature part of $\mathrm{Im}\Pi_T$, namely $\pi_0$
given by (\ref{pi0}) yields the distribution of photons in the
\emph{virtual photon cloud of the vacuum} given by
\begin{equation}\label{vacvirtual}
\frac{dN_{vv}}{d^3xd^3k} = \frac{1}{k}
\int_{k}^{\infty}\frac{\pi_0(\omega,k)}{(\omega+k)^2}~\frac{d\omega}{8\pi^4}
\; .
\end{equation}
This contribution can be extracted by taking $T=0$ and $t_0
\rightarrow -\infty$ in  $\frac{dN(t)}{d^3xd^3k}$, it is linearly
divergent and it clearly must be subtracted since it is not
observable.

ii) The finite temperature contributions from $\pi_{2P}$ and
$\pi_{LD}$ for $k \gg T$ are dominated by the region $\omega \sim
k$ in the frequency integral in eq. (\ref{yieldb}) which is
\emph{finite}. A lengthy but straightforward calculation from
eqs.(\ref{pi2P}) and (\ref{piLD})  yields the following result for
the frequency integral of both these terms for $k\gg T$
\begin{equation}
\frac{d N^{(V)}(t)}{d^3x d^3k}\Bigg|_{av}-\frac{dN_{vv}}{d^3xd^3k}
\simeq \frac{10 \; \alpha_{em} \;\zeta(3)}{32 \;\pi^4} \;
\frac{T^3}{k^3}+ \mathcal{O}\left(\frac{1}{k^5}\right)
\label{speccloud}
\end{equation}
\noindent where `av' refers to the time average of the oscillatory
term and $\zeta(3) = 1.202057\ldots$. The result (\ref{speccloud})
originates in the region $\omega \sim -k$ both in the Landau
damping as well as two particle contribution. This region features
a contribution that is not exponentially suppressed in $k$ for
$k\gg T$.

It is clear that the integral over the momenta $k$ yields a
logarithmically divergent number of photons \emph{in the medium}
and a linearly divergent energy integral. We identify this
temperature dependent term as describing the virtual photon cloud
\emph{in the medium}, which again must be subtracted since it is
unobservable.

The divergence associated with the virtual cloud of the vacuum
given by eq. (\ref{vacvirtual}) as well as the divergences in the
photon number and energy stemming from the eq.(\ref{speccloud})
must both be subtracted from $\frac{d N^{(V)}(t)}{d^3x d^3k}$.

An alternative and illuminating interpretation of the virtual
cloud of the vacuum emerges by noticing that eq.(\ref{vacvirtual})
is related with the vacuum wave function renormalization $Z$ given
by
\begin{equation}\label{wavefunc}
Z = 1 + \mbox{Re} \left. \frac{\partial
\Pi^T_0(\omega,k)}{\partial\omega^2}\right|_{\omega=k} = 1 -
\frac{1}{2\pi \, k } \int_{-\infty}^{+\infty}d\omega' \;
\frac{\mbox{Im} \Pi_T(\omega',k;T=0)}{(k-\omega')^2} \; ,
\end{equation}
where we used the dispersion relations
eqs.(\ref{disprel})-(\ref{sigret}). We see that
eq.(\ref{vacvirtual}) is just the zero temperature wave function
renormalization to orden $ e^2 $.  The vacuum virtual cloud
dresses the bare particles into the physical `in' or `out' states,
the wave function renormalization is simply the overlap of these.

The asymptotic reduction formula (LSZ formulation) requires that

\begin{equation}\label{LSZ}
a_{\bf{k},\lambda}~;~ a^\dagger_{\bf{k},\lambda} \rightarrow
Z^{-\frac{1}{2}}a_{\bf{k},\lambda;out}~;~
Z^{-\frac{1}{2}}a^\dagger_{\bf{k},\lambda;out}
\end{equation}

Thus multiplying the number operator by $Z^{-1}$,  cancels the
vacuum contribution to $\mathcal{O}(e^2)$ thus justifying the
subtraction of the vacuum term. The asymptotic reduction (LSZ)
formalism requires the \emph{zero temperature} wave function
renormalization since the in and out states are the states created
from the physical \emph{vacuum} by the in and out operators.

In the \emph{medium} the extra in-medium contribution to the
virtual cloud dresses the physical particle into a
\emph{quasiparticle}, in this case a
plasmon\cite{brapis,lebellac}, these are not asymptotic states.
Thus the time dependent terms which  are associated with the
virtual cloud at asymptotically long times are actually describing
the dynamics of formation of the quasiparticle in the medium.

While  subtracting the vacuum term corresponds to multiplying by
the inverse of the zero temperature wave function renormalization
according to the LSZ reduction formula and the propagation of a
physical particle in the out state, a similar interpretation for
the medium contribution is not available. The virtual cloud of the
medium dresses the physical particle into a quasiparticle as it
evolves in the medium but is not an asymptotic state. The
subtraction of the in medium virtual cloud cannot be justified on
the basis of asymptotic theory.

Furthermore, there is no unambiguous manner to subtract the
divergent terms at all times, since the oscillatory terms are
finite, and subtracting solely the time independent terms leaves
an expression that becomes negative, features the divergences
described above at the initial time $t=t_0$ but is finite for any
$t\neq t_0$.

Hence, in this section we proceed to  subtract \emph{completely}
the contribution $\frac{d N^{(V)}(t)}{d^3x d^3k}$ and  use the
following definition of the subtracted yield for the analysis that
follows.
\begin{equation}\label{yieldfina}
\frac{d N(t)}{d^3x d^3k}  \equiv \frac{d N^{(T)}(t)}{d^3x
d^3k}=\frac{1}{(2\pi)^3\,k} \int_{0}^{\infty}
\frac{d\omega}{\pi}\,n(\omega)\,
\mathrm{Im}\Pi_{T}(k,\omega)\left[
\frac{1-\cos(\omega-k)(t-t_0)}{(\omega-k)^2}+
\frac{1-\cos(\omega+k)(t-t_0)}{(\omega+k)^2}\right] \; .
\end{equation}
This expression is finite and positive at all time and the large
$\omega, k$ regions are exponentially suppressed.

We emphasize that completely subtracting $\frac{d N^{(V)}(t)}{d^3x
d^3k}$ also neglects the positive and \emph{finite} time dependent
contributions from this term but which cannot be unambiguously
separated from the
 divergent terms during a finite time interval.

At asymptotically long time the definition of the yield
(\ref{yieldfina})  features a term that is constant in time and
terms that grow either linearly or logarithmically (or both) in
time. The time independent and finite contribution which is
typically associated with the virtual photon cloud can  be
separated unambiguously from the photons produced with a constant
rate \emph{only in the asymptotic long time limit}. However, since
the QGP has a finite lifetime, only the virtual cloud of the
vacuum and the medium contribution that leads to a divergent
number of photons and energy can be unambiguously associated with
the virtual photon cloud of the medium. Furthermore, these
(divergent) contributions are associated with very fast
oscillations and become constant in time on a very short time
scale ($\ll 1~\mathrm{fm}/c$) as clearly shown  in fig.
\ref{fig:vircloud}.

To be sure the photons produced by the plasma are detected far
away from the collision region and (practically) at infinite time.
However, these photons had been \emph{produced}  in the plasma
during a (much shorter) time scale $t_0 \leq t \leq t_f$ with
$t_f$ being the hadronization time. Clearly the vacuum part of the
virtual cloud can be recognized and subtracted  unambiguously, it
is given by eq. (\ref{vacvirtual}). However, identifying the
contribution to the virtual photon cloud in the medium  can only
be achieved unambiguously if the formation time of the virtual
cloud is much shorter than the lifetime of the QGP and the number
of photons in the cloud diverges. As mentioned above the time
dependent terms which asymptotically are associated with the
virtual cloud are describing the dynamics of formation of the
\emph{quasiparticle} in the medium.

In section \ref{sec:adiabatic} below we will introduce and
implement a method that allows to separate the divergent
contributions to the virtual cloud which are responsible for the
rapid oscillations, in an effective manner.

The  contribution $\frac{d N^{(T)}(t)}{d^3x d^3k}$ evolves on
longer time scales and contains all of the potentially secular
terms, those that grow linearly, logarithmically etc. Furthermore,
since the large $\omega$ regions are exponentially suppressed, the
integrals are dominated by the region $\omega \sim k$ and $\omega
\sim 0$. The region $\omega \sim k$ leads to secular terms from
the resonant denominators, the non-resonant terms feature
oscillations on the time scale $\sim 1/k$, hence their time
dependence is relevant during the lifetime of the QGP specially
for long wavelengths.  Thus, no subtractions on this contribution
are warranted.

\subsection{Real time vs. S-matrix yields:}

We can now establish a comparison between the yields and spectra
predicted by the real time expression eq.(\ref{yieldfina}) with
those
 obtained from the S-matrix approach (the equilibrium
 \emph{rate}). The  equilibrium \emph{rate} in
leading logarithmic order approximation   in the strong coupling
$\alpha_s$ is given in ref.\cite{AMY}, and for two flavors (up and
down) of quarks (and three colors) becomes
\begin{eqnarray}\label{amyformula}
\frac{d N_{SM}}{d^3k d^4x } & = &  \frac{40 \pi T^2}{9(2\pi)^3} \;
\alpha_{em} \; \alpha_s(T) \;
\frac{n_f(k)}{k}\left[\ln\left(\frac{\sqrt{3}}{4\pi\alpha_s(T)}
\right)+C_{tot}\left(\frac{k}{T}\right)\right] \nonumber\\
C_{tot}(z) & = & \frac{1}{2}\ln(2z)+\frac{0.041}{z}-0.3615+1.01 \;
e^{-1.35\,z}+\sqrt{\frac{4}{3}}
\left[\frac{0.548}{z^{\frac{3}{2}}} \, \ln\left(12.28+\frac{1}{z}
\right)+\frac{0.133 \; z}{\sqrt{1+\frac{z}{16.27}}} \right] \; ,
\end{eqnarray}
\noindent where $n_f(k)$ is the Fermi distribution function. This
fit seems to be very accurate in the region of momenta $0.2 \leq
k/T \leq 50$\cite{AMY}.

We will also use  the lattice parametrization \cite{karsch} for
the temperature dependence of the strong coupling $\alpha_s(T)$
given by
\begin{equation}\label{alfastrongofT}
\alpha_s(T)= \frac{6\pi}{29 \ln\frac{8T}{T_c}}~~;~~T_c \sim
0.16~\mathrm{Gev} \; .
\end{equation}
Although this lattice fit is valid at high temperatures and
certainly not near the hadronization phase transition, we will
assume its validity in the temperature range relevant for RHIC in
order to obtain a numerical estimate of the S-matrix yield. We
note, however, that at $\alpha_s(0.3 ~\mathrm{Gev})\sim 0.24 $ and
the validity of the perturbative expansion is at best
questionable.

The yield per unit phase space as a function of time obtained from
this rate is given by
\begin{equation}\label{eqyield}
\frac{d N_{SM}(t)}{d^3k d^3x }=(t-t_0) \; \frac{d N_{SM}}{d^3k
d^4x} \; .
\end{equation}
We begin by comparing the yield from the real time evolution with
the photon polarization in the hard thermal loop approximation,
valid for $k\ll T$.

Since the hard thermal loop (HTL) limit eq. (\ref{HTL}) is valid
for $k\ll T$ and the formula for the rate (\ref{amyformula}) is
valid for $0.2<k/T<50$ the comparison between the two is reliable
for $k/T \sim 0.2$.

\begin{figure}[htbp]
\epsfig{file=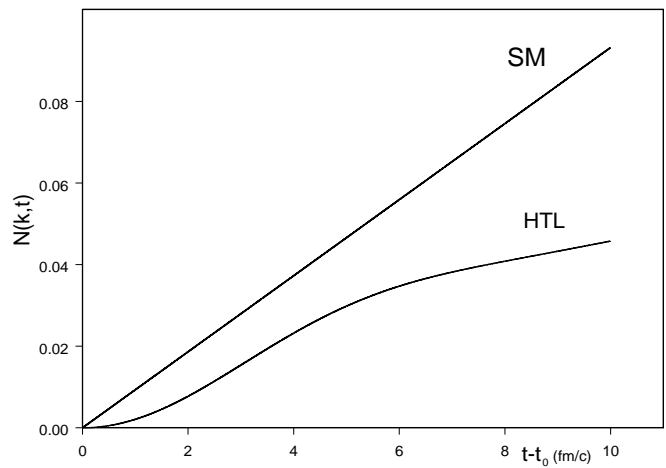,width=4in,height=3in}
 \caption{Comparison between the S-matrix yield SM given by
 eqs.(\ref{eqyield})-(\ref{amyformula}) and the
 real time
 yield (HTL) given by eq.(\ref{yieldHTLoft}) in the HTL
 approximation, both for $k=0.1~\mathrm{Gev};T=0.5~\mathrm{Gev}$ as a
 function
 of $t-t_0$ (in $fm/c$). $N(k,t)=\frac{d^6 N}{d^3k d^3x }$.}
 \label{fig:htlvseq}
\end{figure}

 A comparison of the yields per unit phase space from
the time dependent real time expression (\ref{yieldPi}) given by
eq.(\ref{yieldHTLoft}) in the HTL approximation, and the leading
order result of the S-matrix formulation obtained from
eq.(\ref{amyformula})\cite{AMY}  for $k=0.1~\mathrm{Gev};T=0.5
~\mathrm{Gev}$ is displayed in figure \ref{fig:htlvseq}.

It is clear from this figure that during a finite time interval
compatible with the expected lifetime of QGP in local thermal
equilibrium at RHIC, the  hard thermal loop contribution is of the
same order as the S-matrix result.

\begin{figure}[htbp]
\begin{center}
\epsfig{file=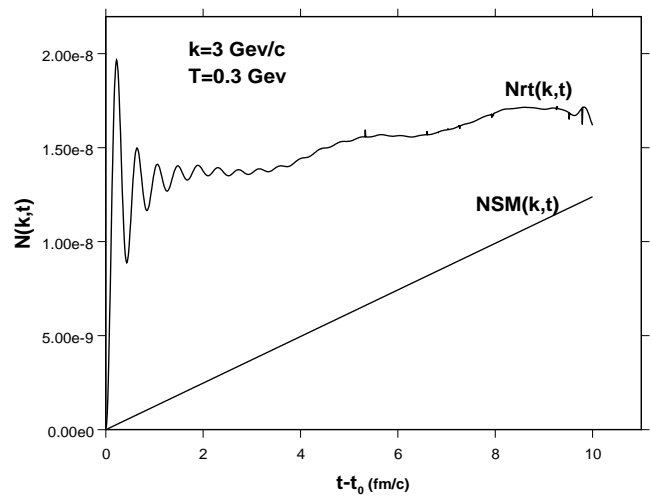,width=4in,height=3in}
 \caption{Comparison between the S-matrix yield $N_{SM}(k,t)$
  given by eqs.(\ref{amyformula})-(\ref{eqyield})
  and the  real time yield   $N_{rt}(k,t)$
  given by eq.(\ref{yieldfina}) for
  $k=3~\mathrm{Gev}/c;T=0.3~\mathrm{Gev}$ as a function
 of $t-t_0$ (in fm/c). The real time yield is dominated by   the
  Landau damping contribution
 given   by eq. (\ref{piLD}). $N(k,t)=\frac{d^6 N}{d^3k d^3x }$.}
  \label{fig:subcompare}
 \end{center}
\end{figure}

Fig. \ref{fig:subcompare} compares the real time yield
eq.(\ref{yieldfina}) with  the full one loop photon polarization
given by eqs. (\ref{totpi}-\ref{piLD}) to the S-matrix yield for
$k=3~\mathrm{Gev}/c\,;\, T=0.3~\mathrm{Gev}$. The real time yield
is dominated by the Landau damping contribution and clearly
competes with the S-matrix result during the lifetime of the QGP
at RHIC.

\begin{figure}[htbp]
\begin{center}
\epsfig{file=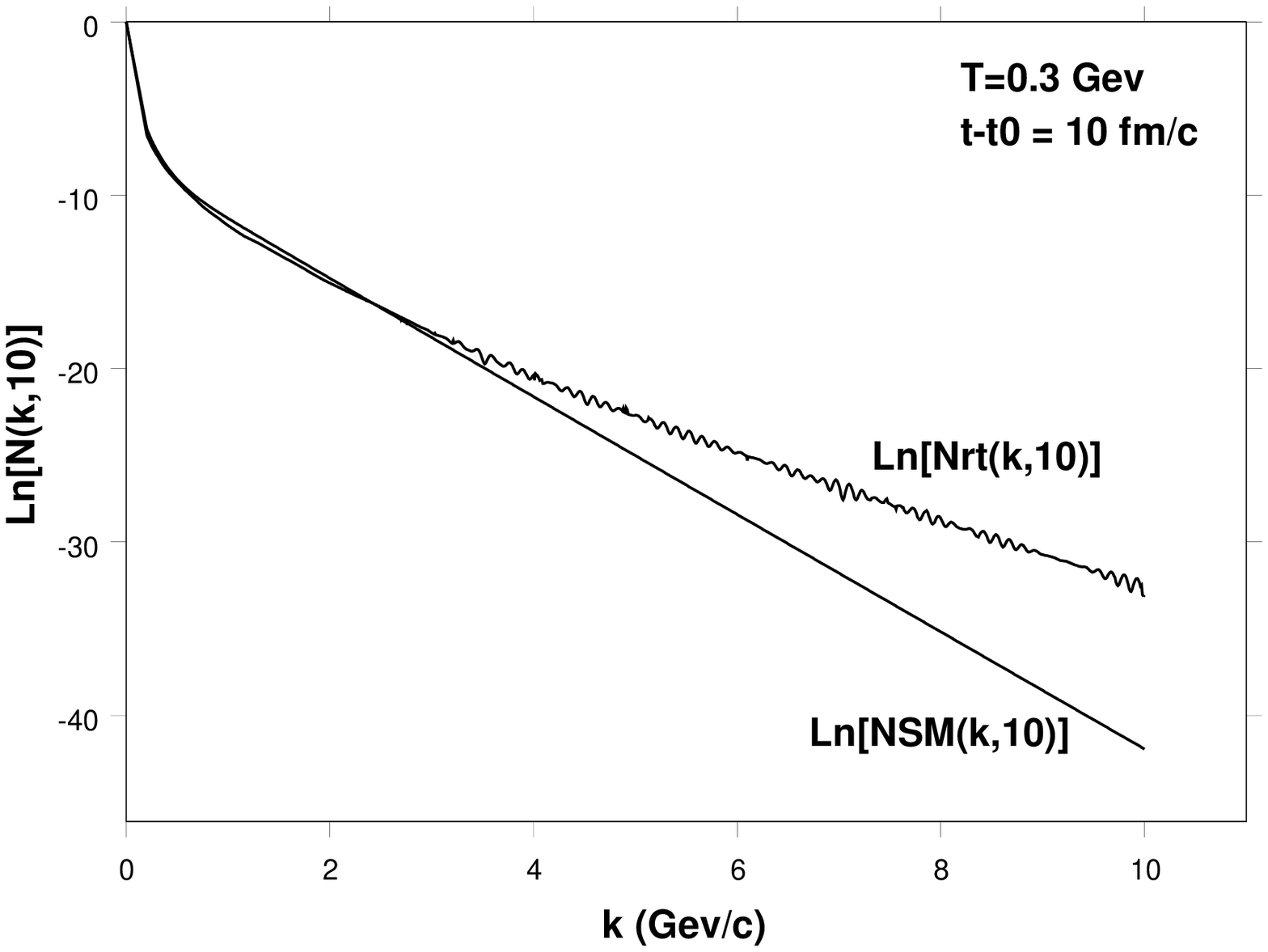,width=4in,height=3in}
 \caption{Comparison of the spectra, $\ln N(k,t) $ vs. $k$
 between the S-matrix $N_{SM}$ and real time $N_{rt}$ yield
 given by eq.(\ref{yieldfina})    for $T=0.3~\mathrm{Gev}$ and
 $t-t_0=10~\mathrm{fm}/c$. } \label{fig:spectrum}
 \end{center}
\end{figure}

Fig. \ref{fig:spectrum} displays the logarithm of  the  real-time
photon yield  given by eq. (\ref{yielda}) vs. $k$ compared to the
S-matrix result for $T=0.3~\mathrm{Gev}\,;\,
t-t_0=10~\mathrm{fm}/c$. Cleary both spectra fall off
exponentially, but the spectrum  from the real time  yield falls
off slower and displays an excess of photons  as compared to the
equilibrium one for $k \geq 2.2 ~\mathrm{Gev/c}$.  Since eq.
(\ref{yielda}) is dominated by the Landau damping contribution to
the imaginary part of the photon polarization, these photons
originate in bremsstrahlung which is a medium effect. We emphasize
that we have only considered the contribution from
eq.(\ref{yielda}) since we have subtracted the full contribution
from eq.(\ref{yieldb}) which is not exponentially suppressed but
falls of as a power law. The subtraction of eq.(\ref{yieldb}) was
motivated by the fact that this contribution features both vacuum
and in-medium divergences associated with the unobservable virtual
photon cloud and the finite contributions cannot be unambiguously
separated during a finite interval of time. We will revisit this
point in section \ref{sec:adiabatic} below where we introduce a
formulation that allows to separate the virtual cloud by allowing
an initial preparation stage.

\subsection{Energetics}

If the thermalized plasma is emitting photons, the energy radiated
away must be drained from the plasma. In this section we study the
different contributions to the energy radiated away with the
escaping photons. The total energy is conserved, namely
\begin{equation}\label{constenergy}
\mathrm{Tr}\left[ \rho(t) H \right]=
\mathrm{Tr}\left[~e^{-iH(t-t_0)} \; \hat{\rho}(t_0) ~
e^{iH(t-t_0)} \; H\right] =
\mathrm{Tr}\left[\hat{\rho}(t_0)~H\right] \; ,
\end{equation}
\noindent where $H$ is the total Hamiltonian (\ref{totih}).
Passing to the interaction picture of $H_0$ in eq.(\ref{totih})
the equation above (\ref{constenergy}) becomes
\begin{equation}\label{Hexpval}
\mathrm{Tr}\left[ \hat{\rho}(t_0)~H\right] = \mathrm{Tr}\left[
\hat{\rho}_{ip}(t_0)  \; U^{-1}(t,t_0)~\left(H_0 + H_I(t) \right)~
U(t,t_0)\right] \; ,
\end{equation}
\noindent where $U(t,t_0)$, $H_I(t)$ and $\hat{\rho}_{ip}(t_0)$
are given by eqs. (\ref{timeevolop}) and (\ref{rhoiniip})
respectively and
 $U^{-1}(t,t_0)=U(t_0,t)$. Assuming translational invariance in a
 (large) volume $V$, the statement of conservation of energy of
 eq. (\ref{Hexpval})  becomes
 \begin{equation}\label{consener}
\Delta\mathcal{E}_{QCD}(t)+ \mathcal{E}_{\gamma}(t)+
\mathcal{E}_I(t) = 0 \; ,
\end{equation}
\noindent where the second term above corresponds to the energy
per unit volume radiated away by the photons produced in the
plasma [see eq. (\ref{photonnumber})] and we have introduced the
following definitions,
 \begin{eqnarray}
 &&\Delta\mathcal{E}_{QCD}(t)  =  \frac{1}{V}~\mathrm{Tr}\left[
 \hat{\rho}_{ip}(t_0) \;  U^{-1}(t,t_0)~H_{QCD}~ U(t,t_0)\right]  -
 \frac{1}{V}~\mathrm{Tr}\left[\hat{\rho}_{ip}(t_0)
~H_{QCD}~\right]  \label{EQCDoft}\; ,\\
&&\mathcal{E}_{\gamma}(t)  =  \int {d^3 k}~ k~
\frac{dN(k,t)}{d^3xd^3k} \label{Egamma} \quad , \quad
\mathcal{E}_I(t)  =  \frac{1}{V}~\mathrm{Tr}\left[
\hat{\rho}_{ip}(t_0) \;  U^{-1}(t,t_0)~H_{I}(t)~ U(t,t_0)\right]
\label{Eintoft}\; .
\end{eqnarray}
The individual terms above can be computed to lowest order in
$\alpha_{em}$ by expanding the time evolution operator up to
second order in the interaction. A lengthy but straighforward
computation using the initial density matrix given by eq.
(\ref{rhoininophots}) and introducing an intermediate set of
states leads to the following result
\begin{equation}\label{terms} \left\{
\begin{array}{c}
 \Delta \mathcal{E}_{QCD}(t) \\
  \mathcal{E}_{\gamma}(t)  \\
 \mathcal{E}_I(t)\\
  \end{array} \right\} =  \int \frac{d^3 k}{(2\pi)^3} \;
  \frac{1}{k} \;  \int_{-\infty}^{+\infty} \frac{d\omega}{\pi}
  \left\{\begin{array}{c}
 -\omega \\   k  \\
 \omega-k \\   \end{array} \right\}
  \; n(\omega)~\mathrm{Im}\Pi_{T}(\omega,k) \;
\frac{1-\cos(\omega-k)(t-t_0)}{(\omega-k)^2} \; ,
\end{equation}
\noindent which manifestly satisfies conservation of energy as in
eq. (\ref{consener}). The asymptotic long time limit is determined
by the region of the spectral density $\omega \sim k$, thus it is
clear from the expressions in eq. (\ref{terms}) that the
interaction energy shuts-off in the long time limit and for
asymptotically long time the rate of radiative energy loss by the
photons is balanced by the rate of energy loss of the plasma,
which can be interpreted as  radiative cooling, namely
\begin{equation}
\frac{d \mathcal{E}_{\gamma}(t)}{dt} = - \frac{d
\mathcal{E}_{QCD}(t)}{dt} \; .
\end{equation}
We emphasize that this result is only valid in the long time
limit, during a finite time interval there is a contribution from
the interaction energy which necessarily is present to satisfy
energy conservation.

As discussed in section \ref{subsec:cloud}  for the virtual photon
cloud, the negative frequency contribution features the zero
temperature divergence associated with the virtual photon cloud of
the vacuum, as well as the  divergence associated with the virtual
photon cloud in the medium. The energy in the photon cloud in the
vacuum diverges as the fourth power of a cutoff, while the energy
in the medium contribution of the virtual cloud diverges linearly
with a cutoff since the number of photons diverges only
logarithmically, as can be seen from eq. (\ref{speccloud}).
Furthermore since the virtual cloud builds up in a very short time
scale, the negative frequency contribution without the
Bose-Einstein distribution averages to a time independent constant
on a time scale $\ll 1~\mathrm{fm}/c$, as can be gleaned from fig.
\ref{fig:vircloud}. Since this contribution features the
divergences associated with the vacuum and in-medium virtual
photon clouds, we subtract it from the energy, consistently with
eq. (\ref{yieldfina}).

Therefore we now study the following subtracted energies,
\begin{eqnarray}\label{posiener} \left\{
\begin{array}{c}
 \Delta \mathcal{E}_{QCD}(t) \\
  \mathcal{E}_{\gamma}(t)  \\
 \mathcal{E}_I(t)\\
  \end{array} \right\} =  \int \frac{d^3 k}{(2\pi)^3}
  \; \frac{1}{k} \; \int_{0}^{\infty} \frac{d\omega}{\pi} \;
    n(\omega)~\mathrm{Im}\Pi_{T}(\omega,k)&&\Bigg[\left\{\begin{array}{c}
 -\omega \\
  k  \\
 \omega-k\\
  \end{array} \right\}
\frac{1-\cos[(\omega-k)(t-t_0)]}{(\omega-k)^2}+ \nonumber \\ && +
\left\{\begin{array}{c}
\omega \\
  k  \\
 -\omega-k\\
  \end{array} \right\}\frac{1-\cos[(\omega+k)(t-t_0)]}{(\omega+k)^2}
 \Bigg] \; .
\end{eqnarray}
\noindent In subtracting the negative frequency contribution
without the Bose-Einstein distribution function, we are also
neglecting the \emph{finite} parts of the negative frequency
contribution to $\mathcal{E}_{\gamma}(t)$, which as mentioned
above cannot be extracted unambiguously.

\begin{figure}[ht]
\begin{center}
\epsfig{file=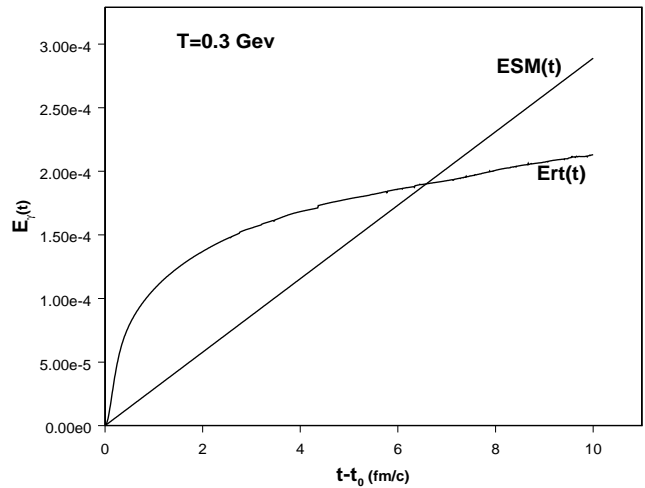,width=4in,height=3in}
 \caption{Comparison between the energy density radiated in
 photons as a function of time for $T=0.3$ Gev. The real time energy
 radiated in photons $E_{rt}(t)$  displays
the logarithmic time dependence. The energy obtained from the
S-matrix yield is $E_{SM}(t)$. } \label{fig:energyphoton}
 \end{center}
\end{figure}

Fig. \ref{fig:energyphoton} displays the subtracted   energy
radiated in photons [see eq. (\ref{posiener})] as a function of
time for $T=0.3~\mathrm{Gev}$ as compared to the energy obtained
from the S-matrix yield. The real-time energy reveals the
logarithmic growth in time and it is of the same order as that
obtained from the S-matrix yield during the lifetime of the QGP
expected at RHIC.

\begin{figure}[ht]
\begin{center}
\epsfig{file=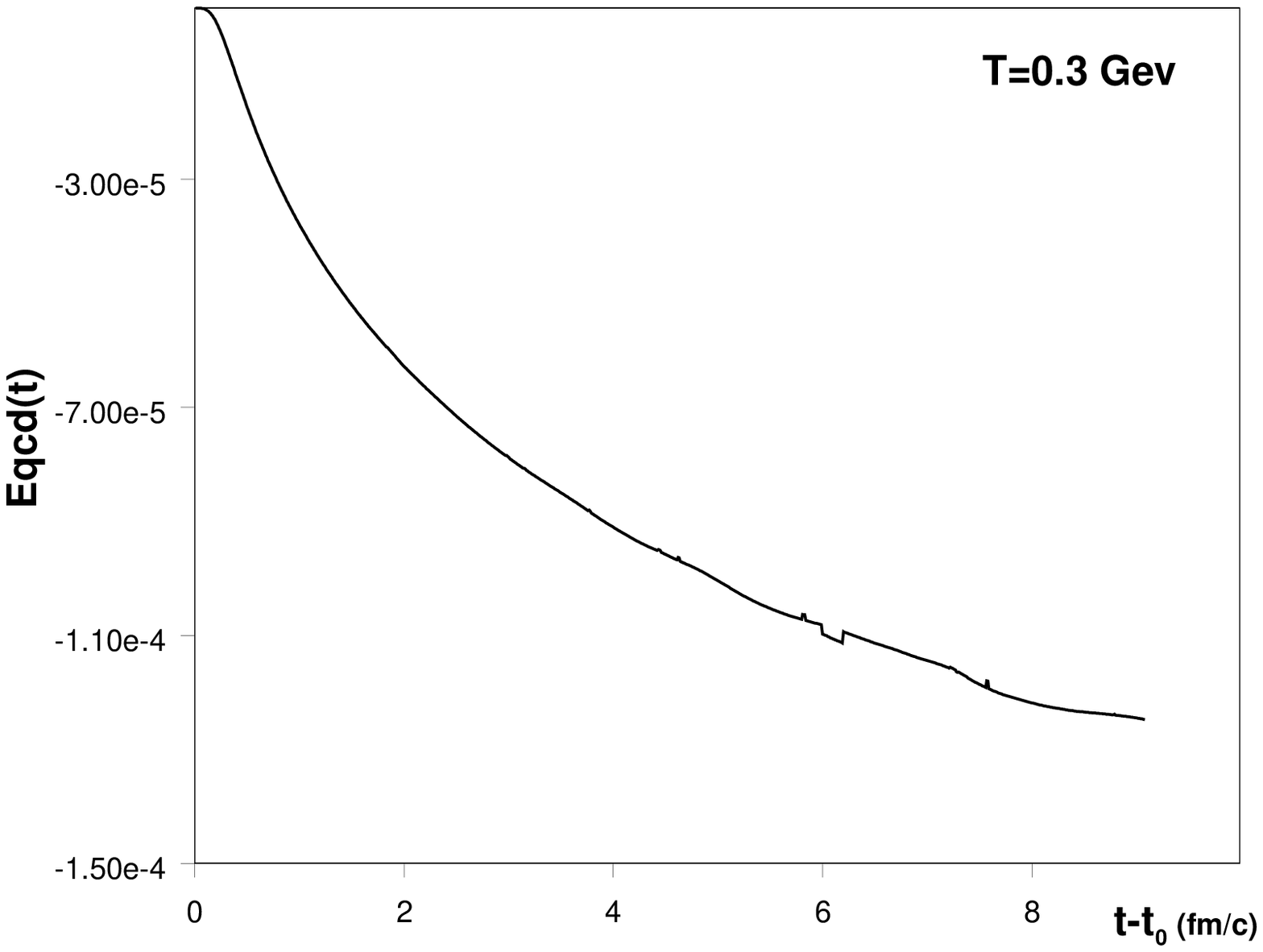,width=3in,height=3in}
\epsfig{file=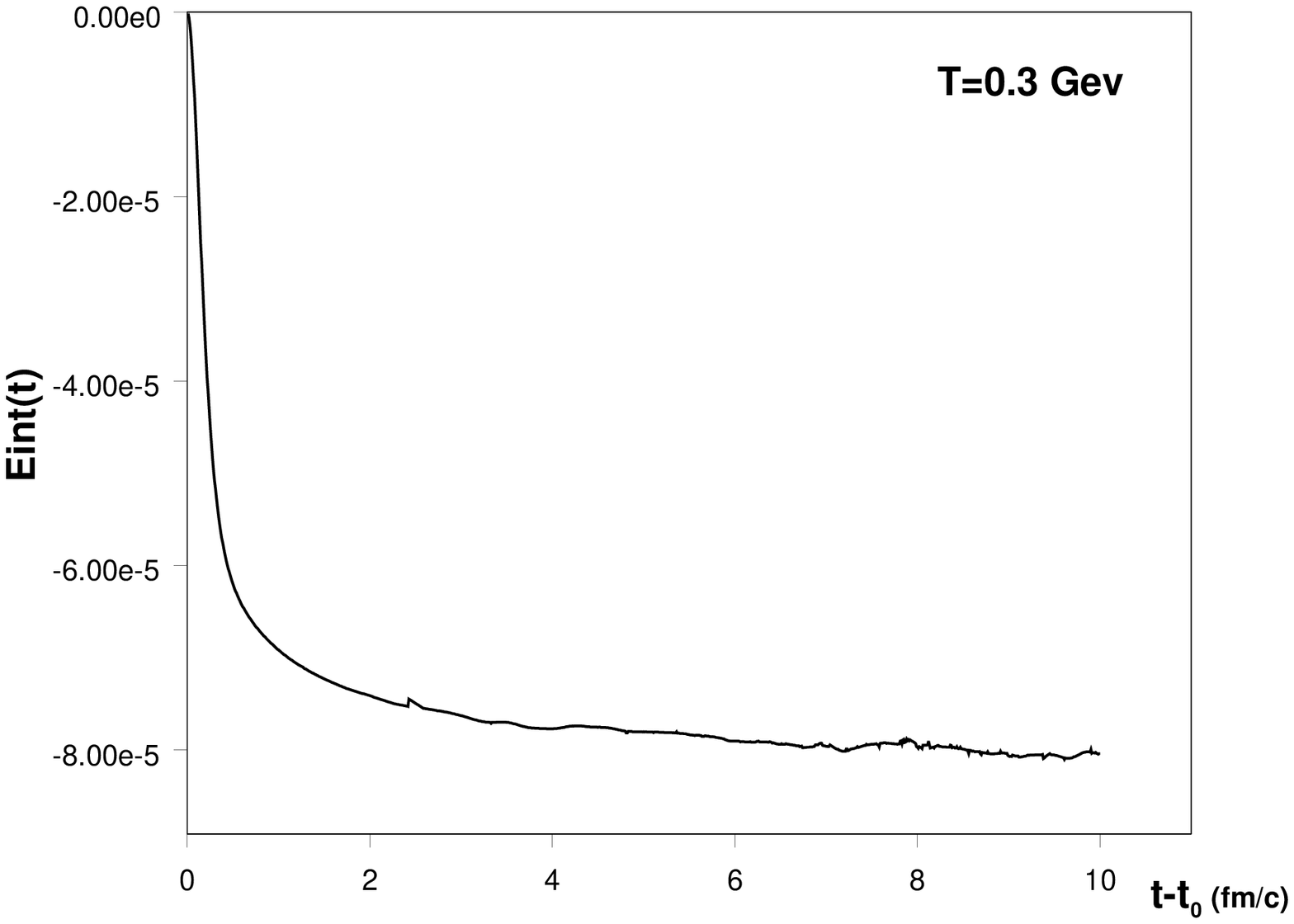,width=3in,height=3in}
 \caption{Positive frequency contribution to $\Delta
 \mathcal{E}_{QCD}(t)$ (left panel) and $\mathcal{E}_I(t)$ (right
 panel)
 vs. $t-t_0$ for $T=0.3~\mathrm{Gev}$.  } \label{fig:energies}
 \end{center}
\end{figure}

Figures \ref{fig:energies} display the subtracted  contributions
eq. (\ref{posiener})  to $\Delta \mathcal{E}_{QCD}(t)$ and
$\mathcal{E}_I(t)$, as a function of $t$ for $T=0.3~\mathrm{Gev}$.

 We confirmed numerically
that the main physical mechanism of radiative energy loss is
Landau damping by studying separately the different contributions
to the photon polarization in the energy.

It is clear from the figure that the interaction term evolves on
time scales of order $\sim 7-8 ~\mathrm{fm}/c$, and that the QGP
cools by photon emission faster than the change in interaction
energy, a clear signal that the photons being emitted are mainly a
consequence of the cooling of the QGP. The numerical analysis
reveals that the energy in photons grows and that in the QGP
diminishes logarithmically, while the interaction energy slowly
approaches a constant over the time interval of the same order as
the lifetime.

\section{ Modelling the initial state: quasi-adiabatic interpolating
states:}\label{sec:adiabatic}

The expression for the total direct photon yield at a given time
$t$ given by eq.(\ref{yieldPi}) was obtained from an initial
density matrix corresponding to a QGP in thermal equilibrium and
the photon vacuum. The assumption on the photon vacuum is in
agreement with the assumption of no photons in the initial state
in the S-matrix calculation. The rational for this choice is that
the photons produced from the pre-equilibrium state leave the
system without buildup of the photon population, and the real time
calculation described above makes explicit this choice of initial
state.

The study of the dynamics of the photon cloud above indicates that
a real time description of photon production- a necessary
treatment to address the finite lifetime of a QGP- must include
the analysis of the virtual photon cloud and a consistent and
systematic separation between the contributions from the virtual
photon cloud and the observed photons.

The S-matrix formulation bypasses any discussion of the dynamics
of the photon cloud by initializing at $t_i \rightarrow -\infty$,
taking the final time $t_f \rightarrow +\infty$ and extracting
only the contributions that lead to a linear time dependence in
the yield. In this manner, all constant contributions, such as
that of the photon cloud, as well as those that grow slower than
linear in time, are neglected. This can also be summarized with
the statement that in the asymptotic long time limit, the yield is
independent of the initial conditions.  This statement requires
that the strong interaction states are in thermal equilibrium from
the initial time $t_i \rightarrow -\infty$. As discussed above,
this does not apply to a QGP produced from a collision and
thermalized at $\sim 1~ \mathrm{fm}/c$ \emph{after} the collision
with a finite lifetime of a few $fm/c$.

The detailed analysis of the dynamics of the virtual photon cloud
highlights the difficulty and ambiguity in trying to separate the
contribution from the virtual photon cloud from the contribution
that grows in time during a finite lifetime. In the vacuum the
virtual cloud relates the physical to the  bare state, thus
subtracting the vacuum virtual cloud amounts to studying the
propagation of physical photons. However, the virtual cloud in the
medium dresses a physical particle into a \emph{quasiparticle},
hence the time dependent terms associated with the in-medium
virtual cloud are actually describing the dynamics of formation of
a plasmon quasiparticle.

In section \ref{subsec:cloud} we analyzed the formation of the
virtual photon cloud, and recognized that the negative frequency
contribution features divergences associated with the virtual
cloud of the vacuum as well as of the medium. Clearly these
divergences are unobservable and must be subtracted, leaving
solely finite contributions to the real time yield.

The subtraction of the divergences associated with the virtual
cloud is ambiguous during a finite time interval and we
\emph{defined} the real time yield  by  eq. (\ref{yieldfina}) to
include only the positive and negative frequency contributions
that are suppressed by the Bose-Einstein distribution function.
Such definition leads to a finite and manifestly positive photon
number density, but is not the only possible definition.

In particular the subtraction (\ref{vacvirtual}) of the virtual
photon cloud of the vacuum, namely the $T=0$ contribution to the
photon yield, entails that the quarks are asymptotic states in the
infinite past and the photon cloud was built up during the
evolution of these asymptotic states from $t_i \rightarrow
-\infty$ up to the collision time. However, this assumption does
not correspond to the actual QGP physics.

Before the collision, quarks and gluons must be described as
partons confined inside the nuclei and in terms of their
distribution functions. Associated with the charged partons there
is a photon cloud with a distribution function that  depends on
the charged parton distribution function in the nuclei.

What happens to this photon cloud during and  after the
collision?, obviously this question is very difficult to address
quantitatively, however the following are some
possibilities\footnote{We thank Yuri Dokshitzer for discussions on
these issues.}.

i) The photon cloud is shaken-off by the collision resulting in a
flash of photons during the pre-equilibrium stage. These photons
have a different origin from those being emitted from
parton-parton annihilation or bremsstrahlung discussed in
ref.\cite{srivageiger}. If the photon cloud is shaken off after
the collision, it will form again during the time between the
collision and the thermalization of the plasma $\sim 1
\mathrm{fm}/c$  because of the electromagnetic interaction.

ii) The partons de-confine during the collision becoming free and
the photon cloud is not modified either by the deconfinement or by
the parton-parton re-scattering that leads to a thermalized QGP.

\medskip

Clearly which of these (or other) possibilities describes the
actual physics of the collision cannot be assessed with the
current level of theoretical understanding.

The importance of these questions cannot be underestimated. If the
lifetime of the QGP were truly infinite, the photon yield at long
times would be insensitive to the initial conditions and the
contribution to the total yield from the initial stage will be of
$\mathcal{O}(1/t)$. However, given the short lifetime of the QGP,
the initial condition not only is important but it bears an
imprint in the spectrum. During the finite lifetime of the plasma
there is no clear and unambiguous separation between the photons
produced at a constant rate, those that are produced at a slower
rate and those that are associated with the virtual photon cloud
in the \emph{asymptotic long time limit}.

Although a detailed understanding of these issues is lacking, we
can provide an approximate description that will model the
essential ingredients. Thus, we now  modify the choice of the
initial density matrix in order to account for a period of
electromagnetic dressing of the strong interaction eigenstates
during the pre-equilibrium stage between the collision and the
onset of thermalization.

 For this purpose, we now revisit the
Gell-Mann-Low theorem\cite{gellmann} that obtains the exact
eigenstates of the total Hamiltonian from the free field \emph{in}
states in terms of the adiabatic M$\o$ller wave operator.

Define the M$\o$ller wave operator
\begin{equation}
U_{\epsilon}(0,-\infty) = {\mathrm
T}\,\exp\left[-i\int_{-\infty}^{0}e^{\epsilon t}
H_{I}(t)dt\right]\label{adiaop} \; ,
\end{equation}
\noindent where $H_I(t)$ is the interaction Hamiltonian in the
interaction picture of $H_0$. The Gell-Mann-Low theorem asserts
that if the states $|n\rangle$ are eigenstates of the Hamiltonian
$H_0$, then the states
\begin{equation}
\widetilde{|n\rangle}= U_{\epsilon}(0,-\infty)|n> \; ,
\end{equation}
\noindent are eigenstates of the full Hamiltonian $H$. The
M$\o$ller wave operator adiabatically dresses the non-interacting
bare state to be the full interacting dressed state during an
infinite period of time. If quarks and gluons were truly
asymptotic states the process of scattering, thermalization and
photon emission would indeed be consistent with the dressing of
the bare states from the infinite past. However, quarks and gluons
are not asymptotic states, furthermore the actual collision
involves bound states of quarks and gluons which are liberated
after the nucleus-nucleus collision but the adiabatic hypothesis
is not suitable to describe the process of formation or break up
of bound states. Namely, the adiabatic hypothesis which is the
basis of the S-matrix approach is not suitable to describe the
dynamics of confinement and deconfinement.

 If partons are freed after the
collision, the dressing process has to take place during the
pre-equilibrium stage either completely, if in the process of the
collision the partons shed their virtual cloud, or partially if
the virtual cloud of photons that partons carried as bound states
is also carried after the collision. This discussion brings to the
fore the difficulty in separating unambiguously the virtual cloud
from the observable photons during a finite lifetime, and
manifestly makes clear the inadequacy of the S-matrix approach to
describe any physical process in a QGP of a finite lifetime.

Given that there is no current understanding of these issues, we
now provide an approximate description of the dressing between the
time of collision and the onset of local thermodynamic
equilibrium.

For this purpose, we define the interaction picture (with respect
to $H_0$) \emph{quasi adiabatic interpolating states} up to lowest
order in the electromagnetic coupling as
\begin{equation}\label{dressedstates}
|n_q;\gamma\rangle= \left[1-i\int_{-\infty}^{t_0}e^{\Gamma(t-t_0)}
\; H_I(t) \; dt + \mathcal{O}(e^2)
\right]|n_q\rangle\otimes|0_{\gamma}\rangle  \; .
\end{equation}
The physical interpretation of these states is that the
electromagnetic interaction dresses the eigenstates $H_0$ during a
time scale $\Gamma^{-1}$. A natural time scale to describe the
pre-equilibrum stage, between the collision and termalization is
about $1~\mathrm{fm}/c$.

 These states interpolate
between the exact \emph{in} states when $\Gamma \rightarrow 0^+$
and the eigenstates of $H_0$ for $\Gamma \rightarrow +\infty$. We
note that these are entangled states in the sense that they are
not simple tensor products of QCD and photon states, therefore
these initial states are correlated.

We can now construct an initial density matrix at time $t_0$ in
terms of these states that is reminiscent of a thermal density
matrix but in terms of the dressed states (\ref{dressedstates}),
namely
\begin{equation}
\hat{\rho}^{\Gamma}_{ip}(t_0) = \sum_{n_q}e^{-\beta E_{n_q}} \;
|n_q;\gamma\rangle \langle n_q;\gamma| \; . \label{rhoinidressed}
\end{equation}
The interpretation of this initial density matrix is that the QCD
eigenstates have been dressed by the electromagnetic interaction
on a time scale $1/\Gamma$ which describes the time between the
collision and the onset of thermalization.

It is important to note that this initial density matrix
\emph{does not} describe a state of thermal equilibrium under the
strong interactions because it \emph{does not} commute with
$H_{QCD}$ since the quark electromagnetic current that enters in
the definition of the dressed states does not commute with
$H_{QCD}$.

We highlight this important point:  \emph{any} initial density
matrix that includes the photon cloud, a result of the
electromagnetic interaction \emph{does not} commute with
$H_{QCD}$, hence it cannot describe a state in thermodynamic
equilibrium under the strong interactions. The only manner to
construct an initial density matrix \emph{with} photons in the
initial state and in thermal equilibrium under the strong
interactions is for this density matrix to be factorized into a
tensor product of a density matrix of pure QCD and a density
matrix of free photons. The density matrix given by eq.
(\ref{rhoininophots}) is one such (the simplest) case.

Once the initial density matrix (\ref{rhoinidressed}) is
specified, its time evolution is completely determined by the full
Hamiltonian and given by eq. (\ref{rhotip}). The evolution of the
number operator in time is therefore given by
(\ref{photonnumberoft}), which upon inserting a complete set of
eigenstates of $H_0$ leads to the following result,
\begin{eqnarray}\label{yieldGamma}
&& \frac{d N^{\Gamma}(t)}{d^3x d^3k} = \frac{1}{(2\pi)^3\,k}
\int_{-\infty}^{+\infty} \frac{d\omega}{2\pi}
\mathrm{Im}\Pi_{T}(k,\omega)~n(\omega)\times \nonumber\\&&
\left[\frac{1}{\Gamma^2+(\omega-k)^2}+\frac{2\Gamma^2}{\Gamma^2+(\omega-k)^2}
\frac{1-\cos(\omega-k)(t-t_0)}{(\omega-k)^2}+\frac{2\Gamma}{\Gamma^2+
(\omega-k)^2}\frac{\sin(\omega-k)(t-t_0)}{(\omega-k)}\right] \; .
\end{eqnarray}
The expression eq.(\ref{yieldGamma}) clearly coincides with
eq.(\ref{yieldPi}) in the $\Gamma \rightarrow +\infty $ limit. In
the opposite limit $\Gamma \rightarrow 0^+$ it becomes
\begin{equation}
\frac{d N^{\Gamma}(t)}{d^3x d^3k}\buildrel{t_i\rightarrow
-\infty}\over = \frac{1}{(2\pi)^3\,k} \left\{
\frac{\mathrm{Im}\Pi_{T}(k,\omega=k)}{e^{\frac{k}{T}}-1} \;
(t-t_i)+\int \frac{d\omega}{\pi}
\frac{\mathrm{Im}\Pi_{T}(k,\omega)}{e^{\frac{\omega}{T}}-
1}\mathcal{P}\frac{1}{(\omega-k)^2} \right\}\label{gamma0}\; .
\end{equation}
Thus, in the limit $\Gamma \rightarrow 0^+$ the yield agrees with
eq.(\ref{longtimeyield}) when the initial time $t_0$ is taken to
$-\infty$ and the photoproduction rate coincides with the result
from the S-matrix calculation, as it must be.

Furthermore, using the identity eq. (\ref{tdelta}), the long time
limit $t-t_0 \rightarrow +\infty$ yields,
\begin{equation}
\frac{d N^{\Gamma}(t)}{d^3x d^3k}\buildrel{t-t_0\rightarrow
-\infty}\over = \frac{1}{(2\pi)^3\,k} \left\{
\frac{\mathrm{Im}\Pi_{T}(k,\omega=k)}{e^{\frac{k}{T}}-1}\left[t-t_0+
\Gamma^{-1}\right]+ \int \frac{d\omega}{\pi} \;
\frac{\mathrm{Im}\Pi_{T}(k,\omega)}{e^{\frac{\omega}{T}}-1} \;
\mathcal{P}\frac{1}{(\omega-k)^2}\right\}\label{gammaasy} \; .
\end{equation}
Since $\Gamma^{-1}$ is the time scale between the collision and
the onset of thermalization, eq. (\ref{gammaasy}) coincides with
the S-matrix calculation in the long time limit from the instant
of the collision. Therefore, this initial preparation is a
physically acceptable description insofar as it reproduces the
asymptotic long time limit.

We note that the eq.(\ref{yieldGamma}) does not vanish at $t=t_0$
because of the first term in the bracket, which is time
independent. The value of the photon yield (\ref{yieldGamma}) at
$t=t_0$, namely the contribution determined by the first term in
the bracket in eq.(\ref{yieldGamma}),  can be interpreted as the
total number of photons (per unit phase space) created during the
time scale $\Gamma^{-1}$. These photons correspond to the virtual
cloud as well as the observable photons emitted during the
pre-equilibrium stage.  This is precisely the physics that the
density matrix in terms of the quasi adiabatic states is meant to
describe.

This interpretation becomes clear in the limit $\Gamma \rightarrow
0$ in which case
\begin{equation}
\frac{1}{\Gamma^2+(\omega-k)^2} \buildrel{\Gamma \rightarrow
0}\over=\pi \; \Gamma^{-1} \; \delta(\omega -k) +
\mathcal{P}\frac{1}{(\omega-k)^2} +{\cal O}(\Gamma) \; ,
\end{equation}
\noindent the principal part leads to the divergences associated
with the virtual cloud and since $\Gamma^{-1}$ is the time scale
of preparation, the term with the delta function gives the real
photons produced during the time scale $\Gamma^{-1}$.

We then  subtract the time independent term (first in the bracket)
in eq. (\ref{yieldGamma}), (which does not contribute to the
rate). We thus  define a photon number that vanishes at the
initial time $t_0$ and that is independent of the initial photon
cloud, namely
\begin{equation}\label{yieldGammasub}
 \frac{d N_S(t)}{d^3x d^3k} = \frac{1}{(2\pi)^3\,k}
\int_{-\infty}^{+\infty}  \frac{d\omega}{\pi} \;
\frac{\mathrm{Im}\Pi_{T}(k,\omega)}{e^{\frac{\omega}{T}}-1}
\frac{\Gamma}{\Gamma^2+(\omega-k)^2} \; \left[\Gamma \;
\frac{1-\cos[(\omega-k)(t-t_0)]}{(\omega-k)^2}+
\frac{\sin[(\omega-k)(t-t_0)]}{\omega-k}\right] \; .
\end{equation}
The interpretation of this  definition is gleaned from the
expression
\begin{equation}\label{integ}
 \frac{d N_S(t)}{d^3x d^3k} = \int^{t}_{t_0}dt' \; \frac{d
 N^{\Gamma}(t')}{d^3xdt' d^3k} \; ,
 \end{equation}
 \noindent with $\frac{d N^{\Gamma}(t)}{d^3x d^3k}$ given by eq.
 (\ref{yieldGamma}). Thus, the subtracted number is obviously the
 photon yield between the time at which the plasma is thermalized
 $t_0$ (after the collision) and the time $t$. Hence, this definition
 neglects the
 virtual photon cloud in the initial state and assumes that the observable photons
 produced during the pre-equilibrium stage leave the plasma.

 The extra powers of $\omega-k$ in the denominator in
 eq. (\ref{yieldGammasub}) render the total yield \emph{finite}.

The subtraction of the time independent term in eq.
(\ref{yieldGamma}) has accounted for the divergent contributions
of the virtual cloud of the vacuum and the medium, leaving a
\emph{finite} result for the time dependent yield.

Therefore, this initial preparation and the subtraction of the
total photon number at the time of thermalization, provide a
possible systematic framework to approximate the physics of the
initial state.

In order to assess the (finite) contribution of the virtual cloud,
it proves convenient again to separate the positive and negative
frequency  contributions. We then obtain
\begin{equation}
\frac{d N_S(t)}{d^3x d^3k}=\frac{d N^{(T)}_S(t)}{d^3x
d^3k}+\frac{d N^{(V)}_S(t)}{d^3x d^3k}
\end{equation}
\noindent with
\begin{eqnarray}
 && \frac{d N^{(T)}_S(t)}{d^3x d^3k}  =   \frac{1}{(2\pi)^3\,k}
\int_{0}^{\infty}
\frac{d\omega}{\pi}~\mathrm{Im}\Pi_{T}(k,\omega)~n(\omega)
~\left[\mathcal{T}^{+}\left[\omega,k,t-t_0\right]+\mathcal{T}^{-}
\left[\omega,k,t-t_0\right]\right]
\label{yieldgamposneg}\\
&&  \frac{d N^{(V)}_S(t)}{d^3x d^3k} = \frac{1}{(2\pi)^3\,k}
\int_{0}^{\infty}
\frac{d\omega}{\pi}~\mathrm{Im}\Pi_{T}(k,\omega)~
\mathcal{T}^{-}\left[\omega,k,t-t_0\right] \; , \label{yieldgam0}
\end{eqnarray}
\noindent in terms of  the functions
\begin{equation}\label{funcs}
\mathcal{T}^{\pm}\left(\omega,k,t-t_0\right) =
\frac{\Gamma}{\Gamma^2+(\omega\mp k)^2} \left[\Gamma \,
\frac{1-\cos[(\omega \mp k)(t-t_0)]}{(\omega \mp
k)^2}+\frac{\sin[(\omega \mp k)(t-t_0)]}{\omega \mp k}\right] \; .
\end{equation}
The term $\frac{d N^{(V)}_S(t)}{d^3x d^3k}$ includes the vacuum
contribution given by $\pi_0(\omega,k)$ in
$\mathrm{Im}\Pi_{T}(k,\omega)$. However,  now this vacuum
contribution is \emph{finite} but of order $\Gamma^2/k^2$ thus
leading  to a divergent number of photons which must be identified
with a vacuum contribution to the virtual cloud and must therefore
be subtracted.

 The remaining term
 \begin{equation}
\frac{d N^{(V)}_{VS}(t)}{d^3x d^3k} = \frac{1}{(2\pi)^3\,k}
\int_{0}^{\infty} \frac{d\omega}{\pi}~
\left[\pi_{2P}(\omega,k)+\pi_{LD}(\omega,k)\right]~
\mathcal{T}^{-}\left[\omega,k,t-t_0\right]\label{subsub}\; ,
 \end{equation}
 \noindent is finite and  only depends on the medium.

 A lengthy but straightforward analysis
 of the time average (neglecting the oscillatory functions) yields the
 following result valid in  the limit  $k\gg T,\Gamma$
\begin{equation}
\frac{1}{(2\pi)^3\,k} \int_{0}^{\infty} \frac{d\omega}{\pi}~
\frac{\Gamma^2\left[\pi_{2P}(\omega,k)+
\pi_{LD}(\omega,k)\right]}{\Gamma^2+(\omega+ k)^2} \simeq \frac{10
\; \alpha_{em} \; \zeta(3)}{32 \; \pi^4} \; \frac{T^3 \;
\Gamma^2}{k^5}\label{gamaav}\; ,
\end{equation}
\noindent plus terms that are exponentially suppressed  for $k\gg
T$. Thus, the total yield, and the energy radiated in photons are
\emph{finite}.

\begin{figure}[htbp]
 \epsfig{file=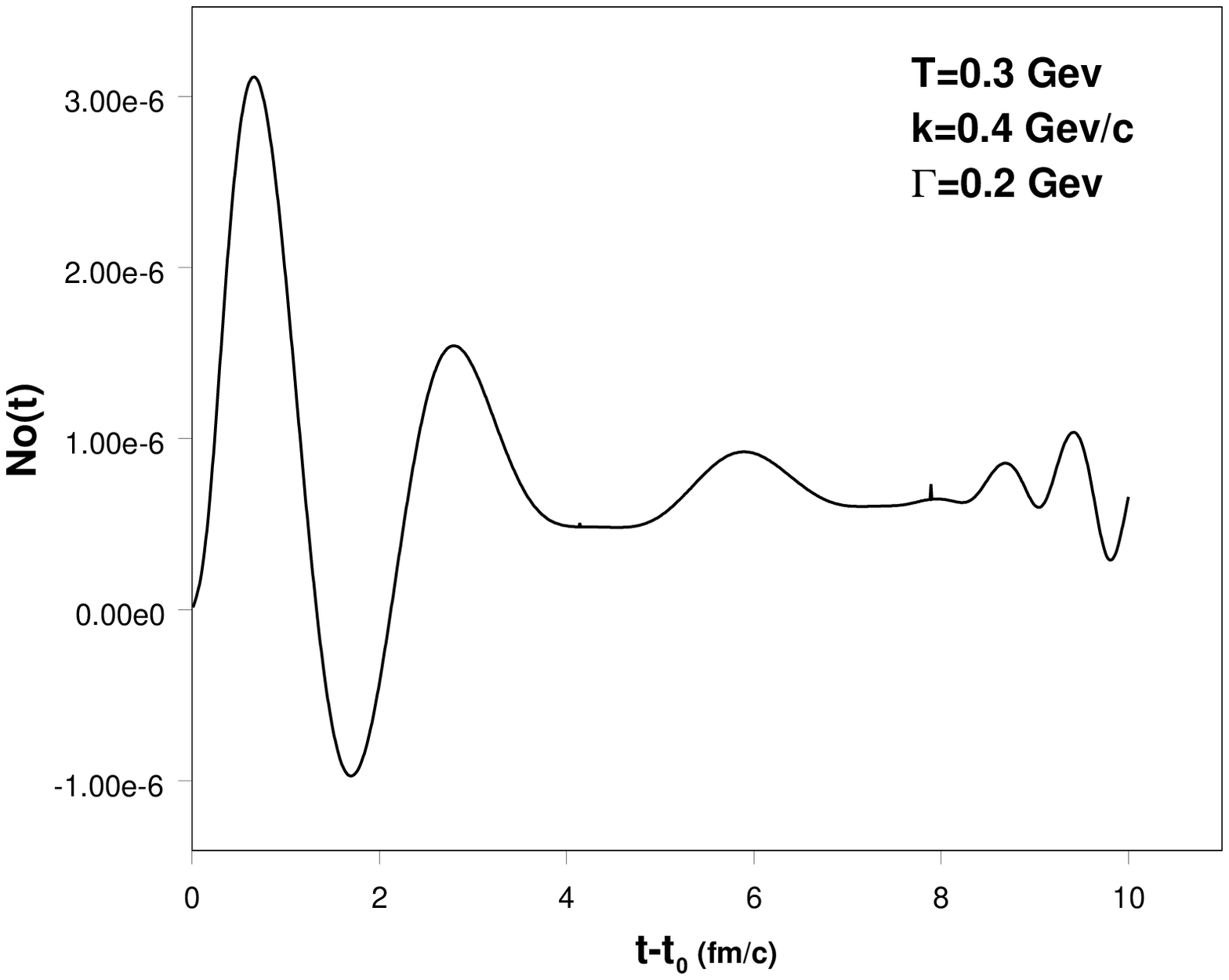,width=3in,height=3in}
\epsfig{file=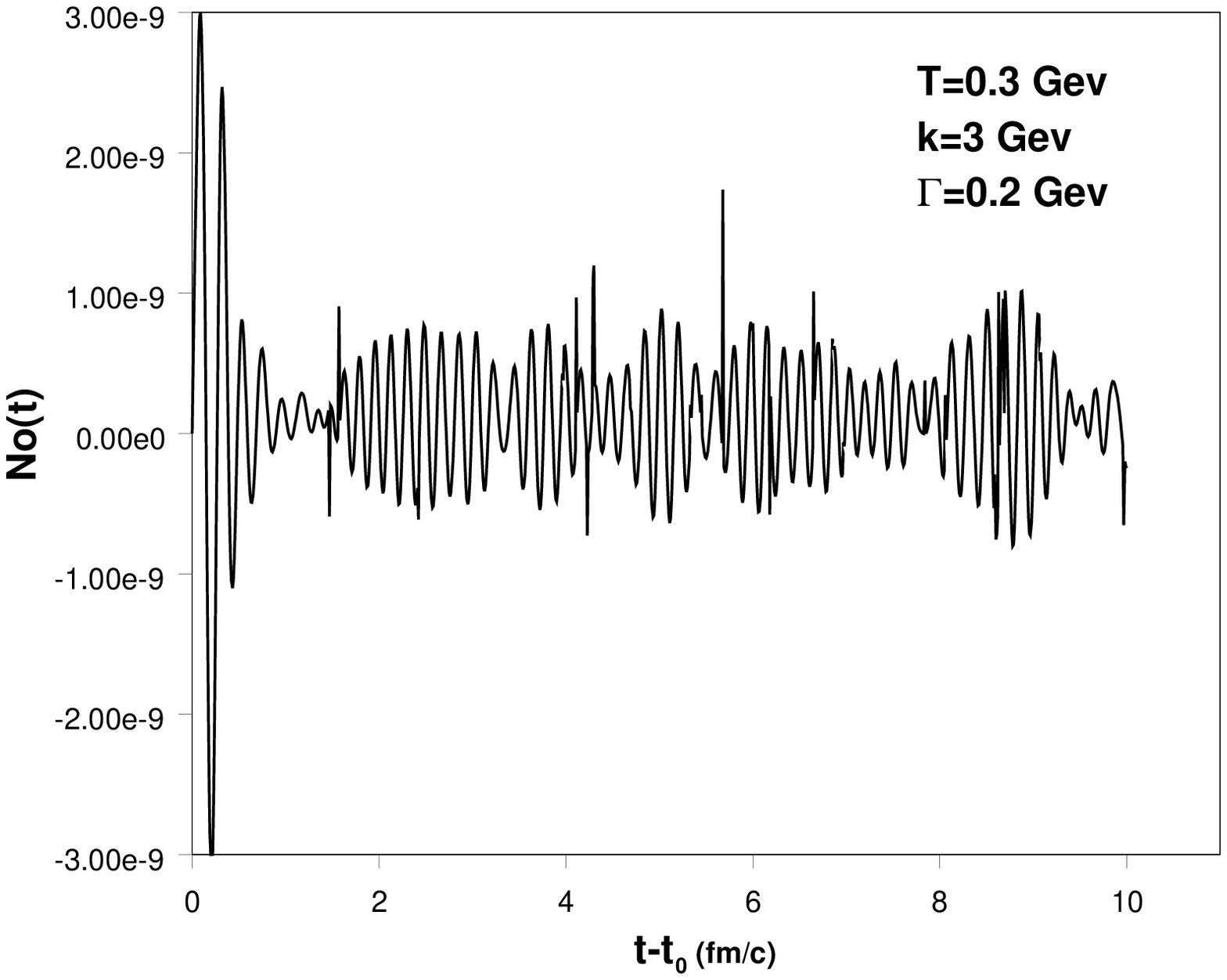,width=3in,height=3in}
 \caption{The contribution of eq. (\ref{subsub}) for
 $k=0.4~\mathrm{Gev}/c$ (left panel) $\,k=3~\mathrm{Gev}/c$ (right
 panel)
 for $\Gamma=0.2~\mathrm{Gev}\,;\,T=0.3~\mathrm{Gev}$.} \label{fig:N0gamma}
\end{figure}

 Fig. \ref{fig:N0gamma} shows  the vacuum subtracted term
$\frac{d N^{(V)}_{VS}(t)}{d^3x d^3k}$ for two values of the
momentum for a range of parameters expected at RHIC.

Comparing fig. \ref{fig:N0gamma} with fig. \ref{fig:vircloud} it
becomes clear  that the oscillations in fig. \ref{fig:N0gamma} are
on much longer  time scales, in particular
 it can be gleaned from fig. \ref{fig:N0gamma} that the
 \emph{shortest} oscillation
 time scale is of $\mathcal{O}(1/k)$ and that there are longer time
 scales. While the
 results leading to fig. \ref{fig:vircloud} are sensitive to the
 cutoffs because the integrals
 diverge, leading to very rapid oscillations,     the frequency
integral in eq. (\ref{subsub})  is finite an the integrand falls
off fast. Hence,  the oscillations are not sensitive to large
frequencies and are on time scales which are of the order of the
lifetime of the QGP.

Thus, the contribution given by eq.(\ref{subsub})  is: i) finite
and leads to a finite photon number and energy, ii) the real time
dynamics is on time scales of the order of the QGP, certainly at
least for $ 0 \leq k \leq 2-3~\mathrm{Gev}/c$. Therefore there is
a priori no reason to subtract this term from the yield and it
must be considered on equal footing as the contribution from eq.
(\ref{yieldgamposneg}).

Thus, the final expression for the photon yield with initial
preparation on a time scale $\Gamma^{-1}$ and after subtracting
the virtual cloud and the pre-equilibrium yield is given by
\begin{eqnarray}
&&\frac{d N_F(t)}{d^3x d^3k} = \frac{1}{(2\pi)^3\,k}
\int_{0}^{\infty}
\frac{d\omega}{\pi}~\Bigg\{\mathrm{Im}\Pi_{T}(k,\omega)~n(\omega)~
\mathcal{T}^{+}\left[\omega,k,t-t_0\right]+ \nonumber\\
&&+\left[\mathrm{Im}\Pi_{T}(k,\omega)~[1+n(\omega)]-
\mathrm{Im}\Pi_{T}(k,\omega;T=0)  \right]\mathcal{T}^{-}
\left[\omega,k,t-t_0\right]\Bigg\} \label{finnumgam}\; ,
\end{eqnarray}
\noindent where $\mathcal{T}^{\pm} \left[\omega,k,t-t_0\right]$
are given by eq. (\ref{funcs}). This expression is one of the
\emph{main} results of this study.

 While we
obtained this expression based on the analysis of the lowest order
contribution, we advocate eq. (\ref{finnumgam}) to lowest order in
$\alpha_{em}$ and \emph{all orders} in $\alpha_s$ as an
\emph{effective} description of the photoproduction yield during a
finite time interval. This expression includes the initial state
preparation and has the following important properties:

\begin{itemize}
\item{The divergences associated with the virtual photon cloud
both in the vacuum and in the medium, as well as the photons
produced during the initial stage prior to thermalization are
subtracted.}

 \item{The total yield as well as the energy are
finite.}

 \item{The limit $t-t_0 \rightarrow +\infty$ correctly
reduces to the photon production rate obtained from the S-matrix
calculation.}

 \item{The limit
$\Gamma \rightarrow 0$ also leads to the S-matrix result for the
rate. This is expected because $\Gamma \rightarrow 0$ corresponds
to an infinitely long preparation stage, which implies an
infinitely long time interval, for which the S-matrix calculation
applies.}

\item{The initial preparation time scale $\Gamma^{-1}$ is a
\emph{parameter} that describes in an effective manner the time
scale between the collision and the onset of the QGP in LTE. It
can be used as a fitting parameter for phenomenological purposes.
}

\end{itemize}

\begin{figure}[htbp]
\epsfig{file=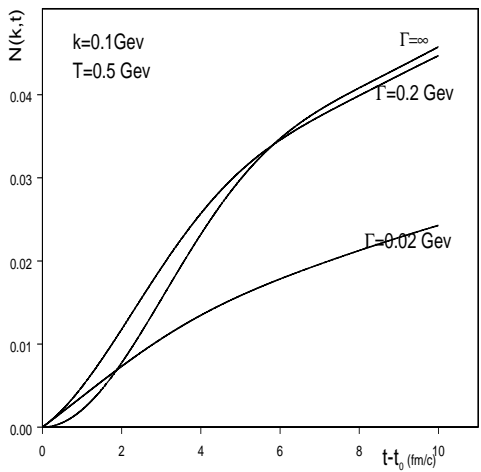,width=3in,height=3in}
 \caption{Comparison between the real time yield with initial
 preparation given by eq.(\ref{finnumgam}) for
 $k=0.1~\mathrm{Gev};T=0.5~\mathrm{Gev}$ as a function
 of $t-t_0$ (in $fm/c$) in the HTL approximation eq. (\ref{HTL}) .}
 \label{fig:iniprep}
\end{figure}

Figure \ref{fig:iniprep}  displays the yield given by
eq.(\ref{finnumgam}) for several values of the initial preparation
time scale for values of $k,T$ for which the HTL approximation is
valid. As mentioned before the HTL approximation gives the leading
contribution for $k\ll T$ and does not depend on the vacuum
contribution.

The values $\Gamma=0.2 ~\mathrm{Gev}$ corresponds to a time scale
of about $1~\mathrm{fm}/c$ which describes the time scale between
the collision and the onset of a thermalized QGP. It is clear from
the figure that change in the yield is rather minor for
long-wavelength photons even  in the case of an extremely long
preparation time scale. Thus for $k\ll T$ the real time yield from
the lowest order $\mathrm{O}(\alpha_{em}\alpha^0_s)$ is of the
same order as the S-matrix yield during the lifetime of the QGP
and is rather insensitive to the preparation time scale.

\begin{figure}[ht]
\epsfig{file=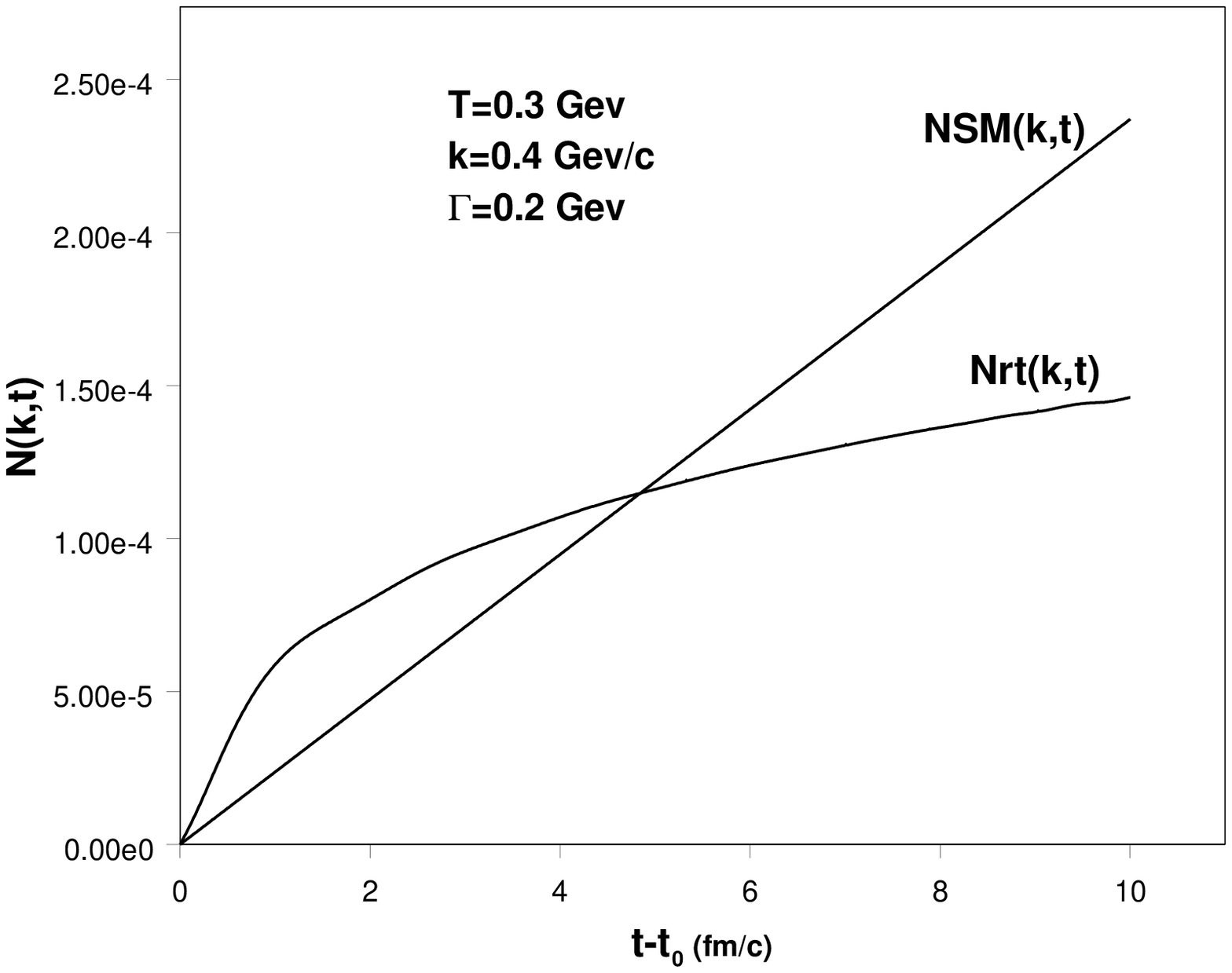,width=3in,height=3in}
\epsfig{file=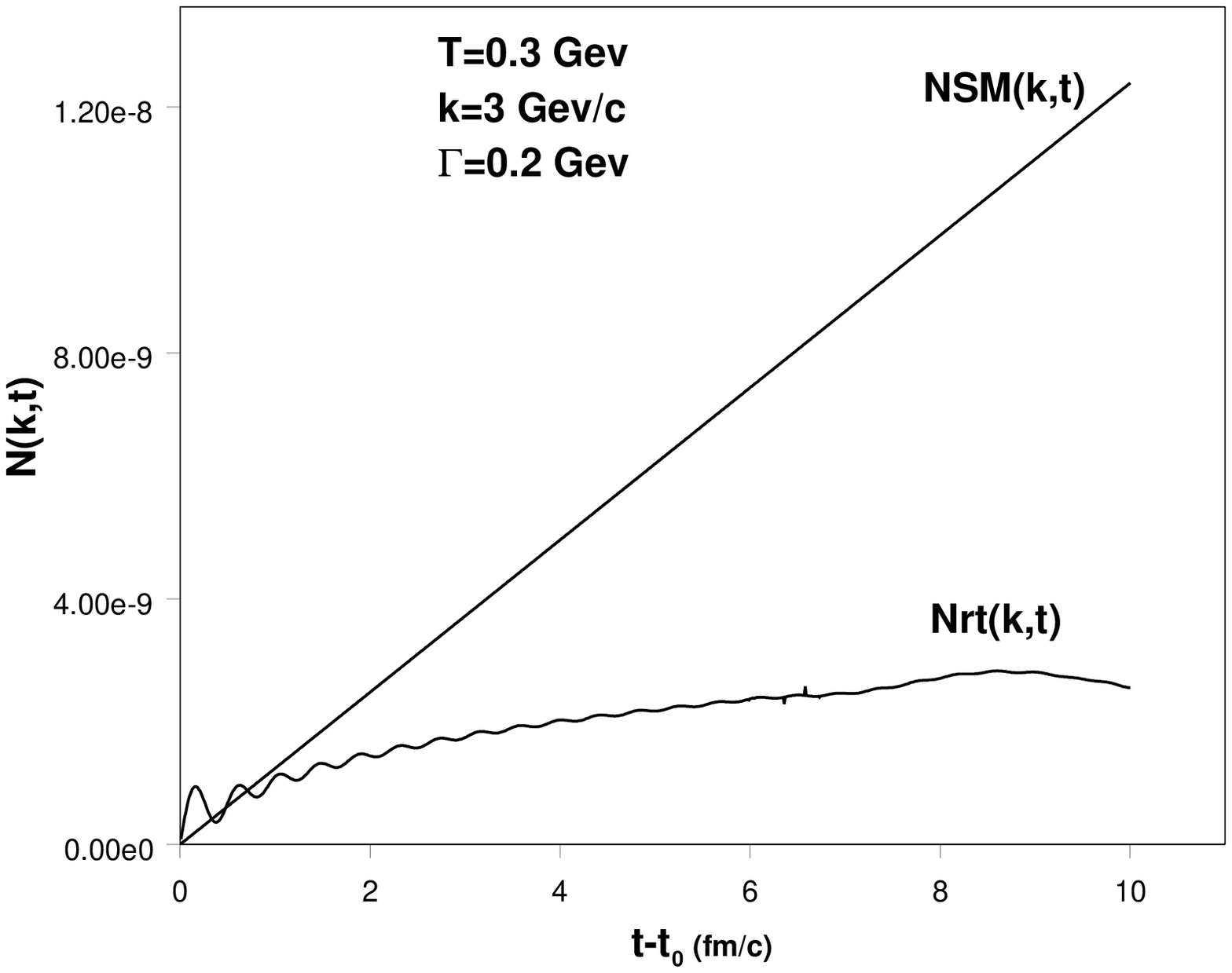,width=3in,height=3in}
 \caption{Comparison between the real time  yield $N_{rt}(t)$  with
 initial preparation given by eqs.(\ref{finnumgam}) with $\Gamma=0.2
 ~\mathrm{Gev}\,;\,T=0.3~\mathrm{Gev}$  and the S-matrix yield as a
 function  of $t-t_0$ (in $fm/c$) for $k=0.4~\mathrm{Gev}$ (left panel)
 and $k=3~\mathrm{Gev}$ (right panel). } \label{fig:numgamma02}
\end{figure}

Figure \ref{fig:numgamma02} shows the comparison between the real
time yield with the full one-loop photon polarization [eqs.
(\ref{totpi}-\ref{piLD})]  and the S-matrix yield as a function of
time for a preparation time scale $1~\mathrm{fm}/c$ for
$k=0.4,3~\mathrm{Gev};T=0.3~\mathrm{Gev}$. Comparing the right
panel of fig. \ref{fig:numgamma02} to the case $\Gamma= \infty$
displayed in fig. \ref{fig:subcompare} we see that they are
qualitatively similar, with the  quantitative difference in the
overall scale. However, it is  clear from these figures  that the
yield from the real time calculation from processes that do not
contribute to the S-matrix rate, is of the same order of or larger
than the yield obtained from the  S-matrix expression during the
lifetime of the QGP.

\begin{figure}[ht]
\epsfig{file=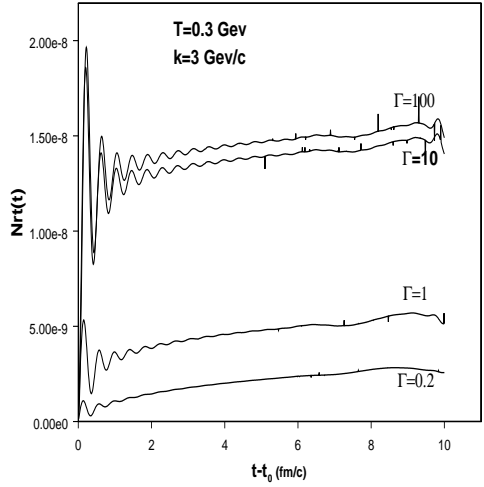,width=3in,height=3in}
 \caption{Comparison between the real time  yield $N_{rt}(t)$  with
 initial preparation given by eqs.(\ref{finnumgam}) for several values
 of  $\Gamma$, for
 $k=3~\mathrm{Gev}/c\,;\,T=0.3~\mathrm{Gev}$. } \label{fig:compara}
\end{figure}

To emphasize this point further for larger values of the momentum,
fig. \ref{fig:compara} displays the real time yield for
$k=3~\mathrm{Gev}/c\,;\,T=0.3 ~\mathrm{Gev}$ for a wide range of
the time  scale for initial preparation.

It is clear from this figure that while there are a few
quantitative changes with respect to the case $\Gamma= \infty$,
qualitatively the results are similar and all of the same order.

\begin{figure}[ht]
\epsfig{file=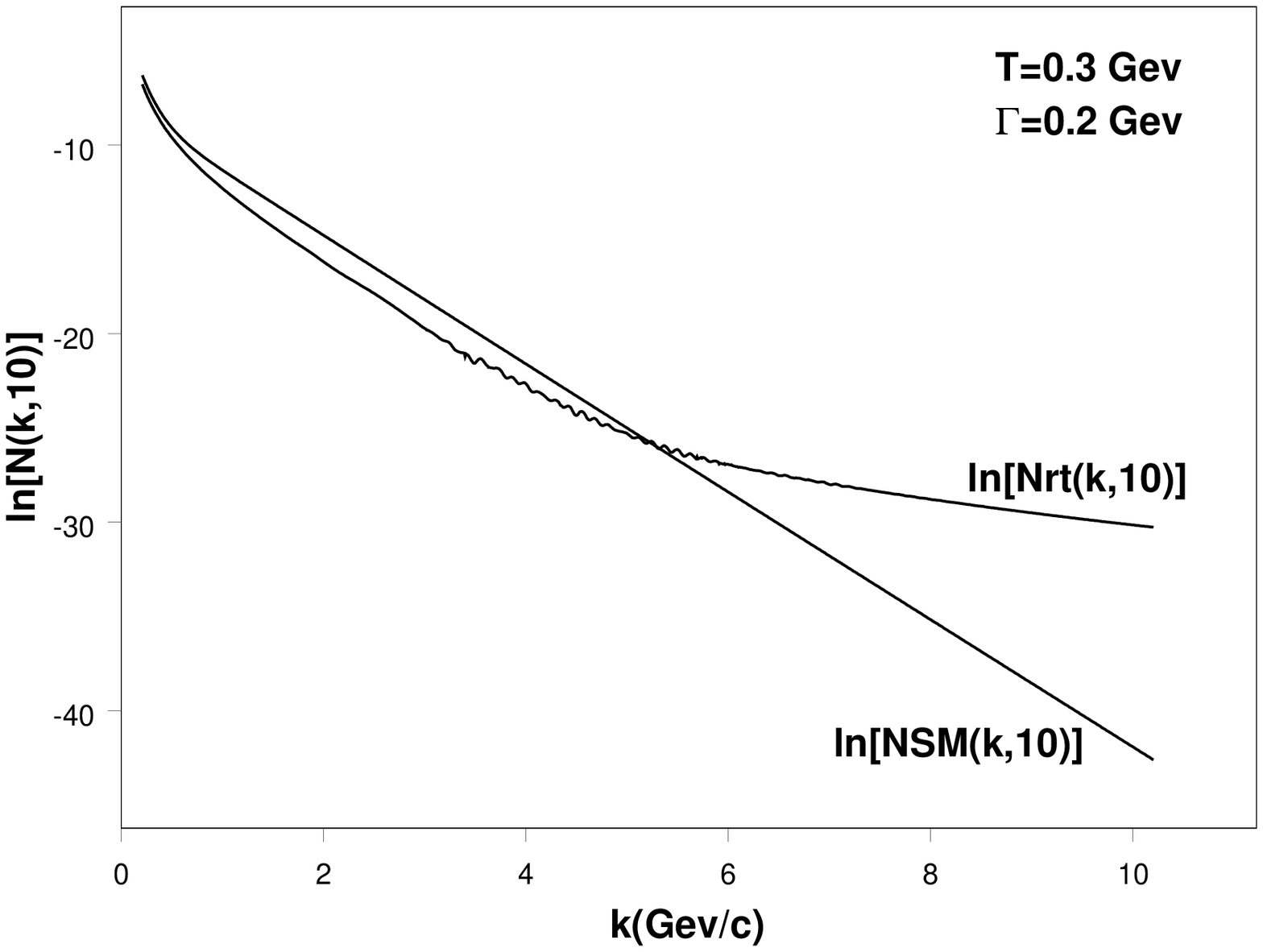,width=3in,height=3in}
\epsfig{file=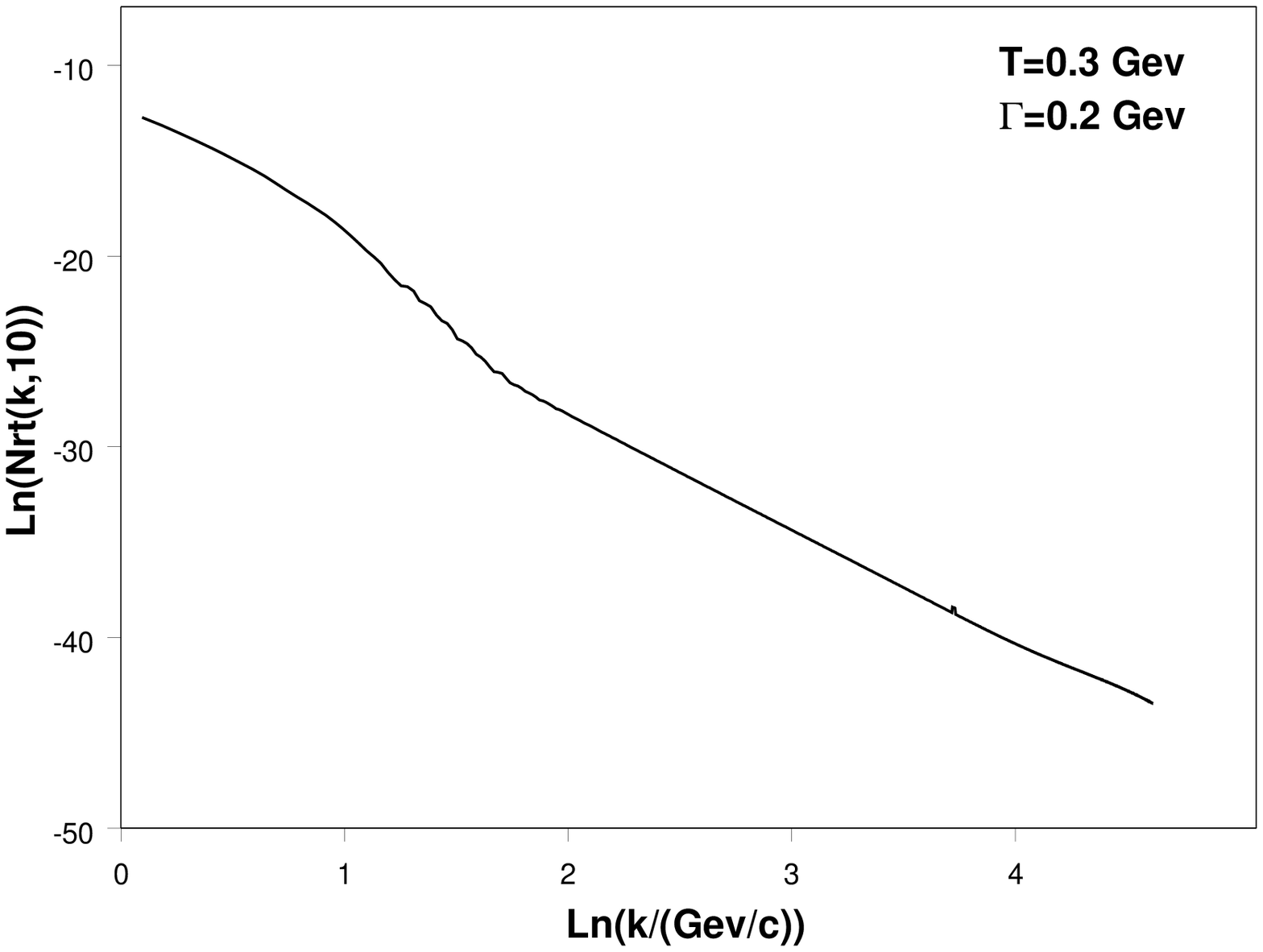,width=3in,height=3in}
 \caption{Comparison between the logarithms of the real time  yield
   $\ln N_{rt}(k,t)$
   with initial preparation given by eqs.(\ref{yieldGammasub}) with
   $\Gamma=0.2 ~\mathrm{Gev}$  and the S-matrix yield
 $\ln N_{SM}(k,t)$ as a function of $k$ (in $Gev/c$) for
 $T=0.3~\mathrm{Gev}$. } \label{fig:lognumgamma02}
\end{figure}

Figure \ref{fig:lognumgamma02} shows the spectrum of photons
produced during the lifetime of the QGP expected at RHIC $\sim
10~\mathrm{fm}/c$ for a preparation time scale of
$1~\mathrm{fm}/c$ and $T=0.3~\mathrm{Gev}$. The left panel
compares the real time and S-matrix yields vs. $k$. The right
panel displays the logarithm of the yield vs. the logarithm of
wavevector for the real time case only and clearly displays the
power law fall of $\sim k^{-5}$ for large momenta ($k\gg
T,\Gamma$) as predicted by eq. (\ref{gamaav}). These figures
suggest a crossover from an exponential to a power law fall in the
spectrum  of the real time yield, the crossover ocurring at a
value $k_c$ which depends on $\Gamma$. We find numerically that
for $\Gamma \sim 0.2~\mathrm{Gev}$ $k_c \approx 2.7-3
~\mathrm{Gev}/c$ resulting  in a marked flattening of the
spectrum. We also find numerically that $k_c$ decreases upon
increasing $\Gamma$. The power law dominance is a telltale  of the
contribution of the term in the real time yield
eq.(\ref{finnumgam}) that does not feature a Bose-Einstein
distribution function which leads to an exponential suppression.
As discussed above this power law leads to a finite number of
photons and a radiated energy. In the infinite time limit the
contribution that leads to this power law would be identified with
the \emph{finite} distribution of photons in the virtual cloud,
but as analyzed and discussed in detail above, during the finite
lifetime this term cannot be separated from the other
contributions and enters in the yield on the same footing.

Thus, the power law spectrum at large momentum is a hallmark of
the processes that contribute during the finite lifetime of the
QGP and that cannot be captured by the S-matrix approach.

We have also studied the energy [see eqs.
(\ref{EQCDoft})-(\ref{Eintoft})], which results in  expressions
similar to those given by eq. (\ref{posiener})  but the positive
 and negative frequency contributions are replaced by those in the
 subtracted yield
  (\ref{finnumgam}). The numerical study of the energy reveals
  minor quantitative changes with respect to the results shown in
  figs. \ref{fig:energyphoton}-\ref{fig:energies}. The
  contributions of the terms which are not exponentially suppressed
  by the Bose Einstein distribution function begin to become
  important when the momentum is of order $k \sim 3-4
  ~\mathrm{Gev}/c$ at which point all contributions are very
  small. The momentum integrals that lead to the energies are
  dominated by momenta $\leq 1.5 -2~\mathrm{Gev}/c$ for $\Gamma
  \sim 0.2~\mathrm{Gev}$. Consequently, the results for the
  energies from the initial density matrix with the initial stage
  preparation are very similar to the results displayed in figs.
   \ref{fig:energyphoton}-\ref{fig:energies} with an overall small
   change in the scale.

\section{Conclusions, discussions and implications}\label{sec:conclusions}

In this article we studied the direct photon production from a QGP
in local thermodynamic equilibrium during a finite lifetime. After
discussing the shortcomings of the usual  approach based on the
S-matrix calculation of the emission \emph{rate}, we focused on
describing photon production directly from the real time evolution
of an initial density matrix.

The main premise of this study is that there are processes that
contribute to the direct photon yield during the finite lifetime
of the QGP but that are \emph{not} captured by the S-matrix
approach. We highlighted this point by restricting our study to
the lowest order contribution of $\mathcal{O}(\alpha_{em})$ to the
yield. While this contribution is subleading in the asymptotic
long time limit (that is if the QGP were truly  a stationary state
of infinite lifetime) it does contribute to the yield during a
finite lifetime.

We began our study by first considering an  initial density matrix
 that  describes a QGP in LTE with no photons in the initial
 state, compatible with all of the assumptions in the literature that lead to
 the S-matrix calculations of the emission rate. This study
 revealed the important aspect of the dynamics of formation and
 build-up of the virtual photon cloud. We highlighted that the finite
 lifetime of the
 transient QGP results in that the photon spectrum retains information
 of the initial state and pointed out the inherent
 ambiguities associated with the separation of the virtual and
 observable photons during the short timescale between formation and
 hadronization.

 While the terms that
 yield divergences in the number of photons and energy from the
 virtual photon cloud of the vacuum and in the medium can be
 identified, there is no unambiguous manner to subtract these from
 the finite contributions during a finite lifetime. Within this
 choice of initial state, we \emph{defined} the yield subtracting
the divergent contributions associated
 with the virtual clouds but also finite, time dependent terms
 because there is no unambiguous manner to extract these. The
 resulting yield clearly shows that there are contributions from
 processes that cannot be captured by the S-matrix approach, but
 that contribute to the direct photon yield during the finite lifetime.

 In particular, within the assumption of no initial photons, our
 study revealed that even after subtracting finite contributions,
 the yield from lowest order processes is of the same order of \emph{or larger}
 than those obtained in the S-matrix calculations.

 We then provided an effective description of the preparation
 of the initial state by constructing the initial density matrix
 of a thermalized QGP in terms of quasi-adiabatic states obtained
 by electromagnetic dressing the QCD eigenstates over a time scale
 $\Gamma^{-1}$. This time scale describes in an effective manner
 the pre-equilibrium stage between the deconfinement of partons
 and the onset of a thermalized QGP, it is expected to be of
 the order of $1~\mathrm{fm}/c$. This quasi adiabatic initial
 condition on the density matrix allows to extract the (divergent)
 contribution from the virtual cloud  built-up as well as the
 observable photons
 emitted during the pre-equilibrium stage. The main result of this
 analysis is eq. (\ref{finnumgam}) which  provides an
 effective but systematic formulation of direct photon production
 from a QGP in local thermal equilibrium directly in real time.
 There are many advantages of this formulation over the usual S-matrix
 approach:

i) it does not suffer from the caveats associated with the
S-matrix approach described in section \ref{sec:Smatrix} ,

ii) it describes photon production in real time as an initial
value problem, consistently with hydrodynamics\cite{boyanphoton},

iii) it includes a description of the initial state in an
effective manner in terms of the parameter $\Gamma$ which is
associated with the inverse of the time scale between the
collision and that of thermalization.

\medskip

This \emph{is} the time scale during which partons are almost free
and parton-parton scattering brings the deconfined partons to a
state of local thermodynamic equilibrium.

The final expression for the real time photon yield given by
eq.(\ref{finnumgam}) has several important properties:

\begin{itemize}
\item{It reproduces the result of the S-matrix result  in the two
limits in which it must be equivalent: the $t-t_0\rightarrow
+\infty$ and the $\Gamma\rightarrow 0$ limits. Both limits
actually refer to a QGP of infinite lifetime, for which the
S-matrix calculation applies. }

\item{It allows to separate the divergent contributions from the
vacuum as well as the in-medium virtual cloud, along with the
observable photons produced during the pre-equilibrium stage.
Therefore this expression leads to a finite photon yield and a
total finite radiated energy. }

\item{This description parametrizes the initial state in terms of
a time scale $\Gamma^{-1}$, which is a phenomenological parameter.
Namely it provides an effective description of the physics between
the time at which nucleus-nucleus collision results in deconfined
quarks and gluons and the time at which a QGP in LTE emerges.  }
\end{itemize}

\textbf{Phenomenological consequences:} The main results of this
study point out that during the finite lifetime of a QGP in LTE
expected at RHIC or LHC there are processes that contribute to the
yield that are not captured by the usual  S-matrix approach. While
we have focused on the lowest order such term, there are already
many important consequences of phenomenological relevance:

\begin{itemize}
\item{The correct direct photon yield is actually \emph{larger}
than that calculated with the S-matrix approach. This is a
consequence of the processes that contribute even at lowest order,
and also processes missed by the S-matrix that arise from the
region of $\omega \neq k$ in the imaginary part of the photon
polarization. Thus, the correct photon yield will be larger than
the current estimates. A reliable estimate of the correction calls
for a re-calculation of the imaginary part of the photon
polarization for all $\omega \neq k$ up to
$\mathcal{O}\left(\alpha_s \ln\frac{1}{\alpha_s} \right)$. Such
calculation is currently not available.}

\item{An important telltale of the processes that contribute to
the yield during a finite lifetime is a \emph{power law} spectrum
of the yield of the form $k^{-5}$. The coefficient of the power
law bears information on the temperature as well as the time scale
for thermalization $\Gamma^{-1}$ of the plasma. This telltale is
in striking contrast with the S-matrix yield which features an
exponential fall-off. Depending on the value of the time scale
$\Gamma^{-1}$ this power law sets in for $k_c \gtrsim
3~\mathrm{Gev}/c$ with $k_c \sim 3~\mathrm{Gev}/c$ for
$\Gamma^{-1} \sim 1~\mathrm{fm}/c$. While this power law spectrum
may be an important signature, it sets in for a region of momenta
in which a description of the QGP in LTE may break down. As
mentioned above, the current data on the elliptic flow parameter
$v_2(p_T)$\cite{jacobs,v2pT} reveals large departures from
hydrodynamics (+pQCD) which relies on a QGP in LTE for $p_T >
2~\mathrm{Gev}/c$. In the region of momentum up to $\sim
2~\mathrm{Gev}/c$ the real time yield is almost indistinguishable
from an exponential fall off, but begins to flatten towards the
power law at about $\sim 3~\mathrm{Gev}/c$. It is possible that
the excess of photons and the flattening of the spectra in the
WA98 data\cite{WA98} may be explained by the processes studied
here and that originate in the finite lifetime of a QGP. }

\end{itemize}

\textbf{More questions:}

\bigskip

Our study indicates that direct photons from a QGP in LTE may not
be a  clean signature of the formation and evolution of the plasma
as originally envisaged. The short transient nature of the QGP
entails that the spectrum carries information on the initial,
pre-equilibrium stage.  However the electromagnetic properties of
the initial state are largely unknown.
 The current estimates of the \emph{rate}
extracted from S-matrix calculations, which assume the existence
of asymptotic states, and infinite QGP lifetime are not completely
reliable, in particular, these are based on a weak coupling
expansion in terms of $\alpha_s$ but at the energy density
conjectured to be achieved at RHIC $\alpha_s \sim 0.24$. Hence the
estimates based on the S-matrix yield are at best qualitative.
During the finite lifetime of the QGP processes that are
completely neglected by the S-matrix approach give contributions
to the yield that are of the same order as those of the
equilibrium calculations, or even larger at large momenta, since
the spectrum from the real time description features a power law
fall off $\sim k^{-5}$ versus the exponential fall off of the
equilibrium yield.

Thus, in order to provide a phenomenologically reliable estimate
the following questions would need to be addressed:

\begin{itemize}
\item{What actually happens to the virtual cloud of photons  after
the collision??, are the virtual photons in the nuclei shaken off
and if so  does this result in a flash of photons during
pre-equilibrium?. }

\item{The real time yield is sensitive to the structure of the
photon polarization for $\omega \neq k$, what is the full
expression of the imaginary part of the photon polarization up to
$\mathcal{O}\left(\alpha_{em}\alpha_s\ln \frac{1}{\alpha_s}
\right)$? }

\item{What is the range in momenta ($k_T$) for which emission from
a hydrodynamically expanding QGP is reliable?, if the data on
elliptic flow for charged particles is extrapolated to photons
(and in principle there is no reason to assume otherwise) then the
local equilibrium description may only be valid up to $k_T \sim
2~\mathrm{Gev}/c$. }

\item{We have discussed above that the terms that are identified
with the virtual photon cloud in the medium asymptotically at long
time are actually providing \emph{dynamical} information on the
formation of the \emph{quasiparticle} in the medium. This is an
aspect that has not been explored before, the formation of a
quasiparticle in the medium does \emph{not} require scattering and
is to lowest order is independent of the mean-free path. This can
be understood simply from the fact that the HTL approximation does
lead to a plasmon quasiparticle but without collisional damping.
Namely, the (transverse) plasmon is a consequence of Landau
damping and not of any on-shell scattering process associated with
a collisional width or a mean free path. As the produced photon
traverses the medium it must necessarily carry with it a
polarization of the medium that dresses the photon into a
quasiparticle. What happens to this induced polarization once the
plasma hadronizes??, is the virtual photon cloud of the QGP
released in a flash during the hadronization transition?.  If so a
power fall off $\sim k^{-3}$ in the spectrum is an unavoidable
consequence of the formation and later dissipation of the virtual
cloud in the medium. }

%new addition # 2
\item{We have studied the consequences of the finite lifetime to
lowest order in the perturbative expansion corresponding to the
one loop polarization. However for $T/T_c$ between $~1-3$, lattice
data clearly shows that the quark-gluon plasma is not free or
weakly interacting. Thus the next step in the program will
consider self-energy and vertex corrections mediated by gluons,
namely to higher order in $\alpha_s$. The strategy will be to
compute the photon polarization including higher order corrections
in $\alpha_s$ and input its imaginary part in the final equation
(\ref{finnumgam}). We expect to report on this study soon.  }
%end of new addition #2

%new addition # 3
\item{As we discussed above in section (\ref{sec:Smatrix}) (see
the discussion under `caveats') the elliptic flow data suggests
the that the hydrodynamic description is not valid for
(transverse) momenta $k_T  > 2~\mathrm{Gev}/c$. Thus the
assumption of LTE upon which all calculations of direct photon
production from a QGP hinge, including the real time formulation
studied in this article, will not be warranted for large momenta.
However, and perhaps more importantly, at large transverse momenta
it is expected that prompt photons produced during the
pre-equilibrium stage during the pQCD parton-parton scattering
will provide a large contribution to the total photon yield. Thus,
as stated in ref.\cite{renk}, the interpretation of the photon
spectrum for large $k_T$ cannot be unambiguous. The current
understanding of prompt photon production during pre-equilibrium
is based on parton cascade calculations\cite{srivageiger} which
invoke a transport description and includes collisions via pQCD
parton scattering cross sections. Such approach implicitly (and
explicitly in the collision term) relies on an S-matrix
description of the parton-parton collisions and is thus subject to
a similar criticism described in section (\ref{sec:Smatrix})
above. Therefore a reliable estimate of the photon yield for large
momenta requires understanding the dynamics beyond LTE and
providing a reliable estimate of prompt photons from the
pre-equilibrium stage. }
%end of new addition #3

\end{itemize}

The experimental importance of electromagnetic probes  of the QGP
warrants a deeper study and assessment of these questions and in
our view a re-evaluation of the current theoretical status on hard
probes.

\acknowledgements The authors thank Yuri Dokshitzer for fruitful
 discussions  and suggestions. We thank E. Mottola, P. Aurenche,
 R. Baier, D. Schiff and B. Mueller for conversations during
initial stages of this work. D. B.  thanks the N.S.F. for partial
support through grants PHY-9988720 and NSF-INT-9905954, and the
hospitality of LPTHE where part of this work was carried out. H.
J. d. V. thanks the Department of Physics and Astronomy at Pitt
for  their warm hospitality.

\appendix

\section{Photon polarization tensor}\label{appendix:piret}

The retarded photon polarization tensor is given by
\begin{equation}
\Pi_{ij,ret}(\vec{x}-\vec{x}',t-t')=
-ie^2\langle\left[J_i(\vec{x},t),J_j(\vec{x}',t') \right]\rangle
\Theta(t-t')\label{Piretar} \; .
\end{equation}
Introducing a complete set of simultaneous eigenstates of
$H_{QCD}$ and the total momentum operator $\vec{P}$ following the
steps described in section \ref{subsec:nophotons} above, we find
\begin{eqnarray}
\langle J_i(\vec{x},t)J_j(\vec{x}',t') \rangle & = & \int
d^3p~d\omega~ e^{-i\vec{p}\cdot(\vec{x}-\vec{x}')+i\omega(t-t')}
\;
\sigma^>_{ij}(\vec{p},\omega) \label{greater} \; , \\
\langle J_j(\vec{x}',t')J_i(\vec{x},t) \rangle & = & \int
d^3p~d\omega ~ e^{-i\vec{p}\cdot(\vec{x}-\vec{x}')+i\omega(t-t')}
\; \sigma^<_{ij}(\vec{p},\omega) \label{lesser} \; ,
\end{eqnarray}
\noindent where $\sigma^>_{ij}(\vec{p},\omega)$ is given by
eq.(\ref{specgreat}) and
\begin{equation}
\sigma^<_{ij}(\vec{p},\omega) =\sum_{n_q,m_q} \; e^{-\beta
E_{m_q}}\langle n_q|{J_i}(\vec{0},0)|m_q\rangle \; \langle
m_q|{J_j}(\vec{0},0)|n_q\rangle \;
\delta^3(\vec{p}-\vec{p}_{n_q}+\vec{p}_{m_q}) \;
\delta(\omega-E_{n_q}+E_{m_q})= e^{\beta \omega} \;
\sigma^>_{ij}(\vec{p},\omega)\label{specless} \; .
\end{equation}
Introducing the Fourier representation of $\Theta(t-t')$ we find
the photon polarization tensor to be given by
\begin{eqnarray}
\Pi_{ij,ret}(\vec{x}-\vec{x}',t-t') & = &  \int
\frac{d^3p}{(2\pi)^3} \; \frac{dp_0}{2\pi} \;
e^{-i\vec{p}\cdot(\vec{x}-\vec{x}')}
 \; e^{ip_0(t-t')} \; \Pi_{ij,ret}(\vec{p},p_0)\label{PiFou} \; , \\
\Pi_{ij,ret}(\vec{p},p_0) & = & (2\pi)^3 \int_{-\infty}^{+\infty}
d\omega \; \frac{\sigma^>_{ij}(\vec{p},\omega)\left[1-e^{\beta
\omega} \right]}{\omega-p_0+i0}~~.  \label{disprel}
\end{eqnarray}
Therefore the result,
\begin{equation}
(2\pi)^3\sigma^>_{ij}(\vec{p},p_0)=
\frac{1}{\pi}\frac{\mathrm{Im}\Pi_{ij,ret}(\vec{p},p_0)}{e^{\beta
p_0}-1}\label{sigret} \; .
\end{equation}
We note that $\mathrm{Im}\Pi_{ij,ret}(\vec{p},p_0)$ is an odd
function of $p_0$ with $\mathrm{Im}\Pi_{ij,ret}(\vec{p},p_0>0)>0$
therefore $\sigma^>_{ij}(\vec{p},p_0)>0$ for \emph{all} values of
$p_0$.

%%%%%%%%%%%%% bibliography

%\input{qgpphotonsbiblio}


\begin{thebibliography}{99}

\bibitem{feinberg}
E.L. Feinberg, Nuovo Cim. {\bf 34A}, 391 (1976), E. Shuryak, Phys.
Lett. B {\bf 78}, 150 (1978); B. Sinha, {\bf 128}, 91 (1983).
\bibitem{mclerran}
L.D. McLerran and T. Toimela, Phys. Rev. D {\bf 31}, 545 (1985).

\bibitem{kapusta}
J.I. Kapusta, P. Lichard, and D. Seibert, Phys. Rev. D {\bf 44},
2774 (1991); {\bf 47}, 4171 (1993).

\bibitem{gale}
C. Gale and J.I. Kapusta, Nucl. Phys. {\bf B357}, 65 (1991).

\bibitem{baier}
R. Baier, H. Nakkagawa, A. Ni\'egawa, and K. Redlich, Z. Phys. C
{\bf 53}, 433 (1992).

\bibitem{ruuskanen}
P.V. Ruuskanen, in {\em Particle Production in Highly Excited
Matter}, NATO ASI Series, Series B: Physics Vol. 303, edited by
H.H. Gutbrod and J. Rafelski, Plenum Press, New York, 1992.

\bibitem{aurenche}
P. Aurenche, F. Gelis, R. Kobes, and H. Zaraket, Phys. Rev. D {\bf
58}, 085003 (1998).

\bibitem{AMY} P. Arnold, G. D. Moore and L. G. Yaffe,
hep-ph/0111107 (2001).

\bibitem{renk} T. Renk, hep-ph/0301133 (2003).

\bibitem{alam}
J. Alam, S. Sarkar, T. Hatsuda, T.K. Nayak, and B. Sinha, Phys.
Rev. C {\bf 63}, 021901 (2001); J. Alam, S. Sarkar, P. Roy, T.
Hatsuda, and B. Sinha, Ann. Phys. \textbf{286}, 159 (2001).

\bibitem{revphoton} T. Peitzmann and M. H. Thoma, Phys. Rep. 364 (2002)
175.

\bibitem{revphoton2} F. D. Steffen, nucl-th/9909035 (1999), Diploma
Thesis,  F. D. Steffen and M. Thoma,  Phys.Lett. B510 (2001), 98.

% observation of direct photon production in Pb+Pb collisions
\bibitem{WA98}
WA98 Collaboration, M.M. Aggarwal {\it et al}., Phys. Rev. Lett.
{\bf 85}, 3595 (2000); nucl-ex/0006007.

% Quark-gluon plasma: general reference

\bibitem{jacobs} P. Jacobs \textit{Measurements of High Density Matter at
RHIC}, Talk presented at the 2002 Slac Summer Institute Topical
Conference, hep-ex/0211031.

\bibitem{geiger}
K. Geiger, Phys. Rep. {\bf 258}, 237 (1995); X.-N. Wang, Phys.
Rep.  {\bf 280}, 287 (1997).

\bibitem{FOPI}
FOPI Collaboration, F. Rami {\it et al.}, Phys. Rev. Lett. {\bf
84}, 1120 (2000).

\bibitem{wong} C. Y. Wong, Phys. Rev. {\bf C 48}, 902 (1993); M.
G.H. Mostafa and C. Y. Wong, Phys. Rev. {\bf C 51}, 2135 (1995).

\bibitem{sarkar} S. Sarkar {\it et al.} J. Phys. G: Nucl. Part.
Phys. {\bf 22} 951 (1996).

\bibitem{ruuskanen2}  D. Anchishkin, V. Khryapa, V. Ruuskanen,
{\it Thermal Dilepton Radiation from Finite Fireball},
hep-ph/0210346.

\bibitem{boyanphoton}
D. Boyanovsky and H.J. de Vega, Phys. Rev. D {\bf 59}, 105019
(1999). D. Boyanovsky, H.J. de Vega, S.-Y. Wang,  Phys.Rev. {\bf
D61}  065006 (2000); S.-Y. Wang, D. Boyanovsky, H. J. de Vega and
D.-S. Lee, Phys.Rev. {\bf D62}
 105026 (2000);
S.-Y. Wang and D. Boyanovsky, Phys. Rev. D {\bf 63}, 051702
(2001); Nucl. Phys. {\bf A 699}, 819 (2002);

\bibitem{bjorken}
J.D. Bjorken, Phys. Rev. D {\bf 27}, 140 (1983).

\bibitem{blaizot}
J.-P. Blaizot and J.-Y. Ollitrault, in {\it Quark-Gluon Plasma~1},
edited by R.C. Hwa, World Scientific, Singapore, 1990.

\bibitem{v2pT} C. Adler \textit{et. al.}, Star Collaboration,
Phys.Rev.Lett. {\bf 90} 032301 (2003).

\bibitem{srivageiger}
D.K. Srivastava and K. Geiger, Phys. Rev. C {\bf 58}, 1734 (1998).

\bibitem{sollfrank}
J. Sollfrank, P. Huovinen, M. Kataja, P.V. Ruuskanen, M. Prakash,
and R. Venugopalan, Phys. Rev. C {\bf 55}, 392 (1997).

\bibitem{srivastava}
D.K. Srivastava and B. Sinha, Phys. Rev. Lett. {\bf 73}, 2421
(1994); nucl-th/0006018; D.K. Srivastava, Eur. Phys. J. C {\bf
10}, 487 (1999).

\bibitem{boyankinetic} D. Boyanovsky and H. J. de Vega,
\textit{Dynamical renormalization group approach to relaxation in
quantum field theory} hep-ph/0302055 (2003), to appear in Ann.
Phys.

\bibitem{brapis}
E. Braaten and R.D. Pisarski, Nucl. Phys. {\bf B337}, 569 (1990);
{\bf B339}, 310 (1990); R.D. Pisarski, Physica A {\bf 158}, 146
(1989); Phys. Rev. Lett. {\bf 63}, 1129 (1989); Nucl. Phys. {\bf
A525}, 175 (1991).

\bibitem{lebellac}
M. Le Bellac, {\it Thermal Field Theory}, Cambridge University
Press, Cambridge, England, 1996.

\bibitem{karsch}
F. Karsch, Z. Phys. C {\bf 38}, 147 (1988).


\bibitem{gellmann} M. Gell Mann and F. Low, Phys. Rev. {\bf 84},
350  (1951).

%%%%%%%%%%%%%%%%%%%%%%


\end{thebibliography}
\end{document}